# Auctions with Severely Bounded Communication

**Liad Blumrosen**          LIADBL@MICROSOFT.COM
*Microsoft Research*
*1065 La Avenida*
*Mountain View, CA 94043*
**Noam Nisan**          NOAM.NISAN@GMAIL.COM
*School of Computer Science and Engineering*
*The Hebrew University*
*Jerusalem 91904, Israel*
**Ilya Segal**          ILYA.SEGAL@STANFORD.EDU
*Department of Economics*
*Stanford University*
*Stanford, CA 94305*

## Abstract

We study auctions with severe bounds on the communication allowed: each bidder may only transmit $t$ bits of information to the auctioneer. We consider both welfare- and profit-maximizing auctions under this communication restriction. For both measures, we determine the optimal auction and show that the loss incurred relative to unconstrained auctions is mild. We prove non-surprising properties of these kinds of auctions, e.g., that in optimal mechanisms bidders simply report the interval in which their valuation lies in, as well as some surprising properties, e.g., that asymmetric auctions are better than symmetric ones and that multi-round auctions reduce the communication complexity only by a linear factor.

## 1. Introduction

Recent years have seen the emergence of the Internet as a platform of multifaceted economic interaction, from the technical level of computer communication, routing, storage, and computing, to the level of electronic commerce in its many forms. Studying such interactions raises new questions in economics that have to do with the necessity of taking computational considerations into account. This paper deals with one such question: how to design auctions optimally when we are restricted to use a very small amount of communication.

This paper studies the effect of severely restricting the amount of communication allowed in a single-item auction. Each bidder privately knows his real-valued willingness to pay for the item, but is only allowed to send $k$ possible messages to the auctioneer, who must then allocate the item and determine the price on the basis of the messages received. (For example, a bidder may only be able to send $t$ bits of information, in which case $k = 2^t$). The simplest case is $k = 2$, i.e., each bidder sends a single bit of information. This is in contrast to the usual auction design formulation, in which bidders communicate real numbers.

While communicating a real number may not seem excessively burdensome, there are several motivations for studying auctions with such severe restrictions on the communication. First, if auctions are to be used for allocating low-level computing resources, they should use only a very small amount of computational effort. For example, an auction for routing a single packet on the Internet must require very little communication overhead, certainly not a whole real number. Ideally, one would like to "waste" only a bit or two on the bidding information, perhaps "piggy-backing" on some unused bits in the packet header of existing networking protocols (such as IP or TCP). Second, the amount of communication also measures the extent of information revelation by the bidders. Usually, bidders will be reluctant to reveal their exact private data (e.g., Rothkopf, Teisberg, & Kahn, 1990). This work studies the tradeoff between the amount of revealed data and the optimality of the auctions. We show that auctions can be close to optimal even using a single Yes/No question per each bidder. Our results can also be applied to various environments where there is a need for discretize the bidding procedure; one example is determining the optimal bid increment in English auctions (Harstad & Rothkopf, 1994). Finally, a restriction on communication may sometimes be viewed as a proxy for other simplicity considerations, such as simple user interface or small number of possible payments to facilitate their electronic handling. A recent paper (Blumrosen & Feldman, 2006) shows that the ideas illustrated in this work extend to general mechanism-design frameworks where the requirement for a small number of "actions" per each player are natural and intuitive.

We examine the effect of severe communication bounds on both the problem of maximizing social welfare and that of maximizing the seller's expected profits (the latter under the restrictions of Bayesian incentive compatibility and interim individual rationality of the bidders, and under a standard regularity condition on the distribution of bidders' valuations). We study both simultaneous mechanisms, in which the bidders send their bids without observing any actions of the other bidders, and sequential mechanisms where messages may depend on previous messages. We find that single-item auctions may be very close to fully optimal despite the severe communication constraints. This is in contrast to *combinatorial* auctions, in which exact or even approximate efficiency is known to require an exponential amount of communication in the number of goods (Nisan & Segal, 2006).[1]

Both for welfare maximization and revenue maximization, we show that the optimal 2-bidder auction takes the simple form of a "priority game" in which the player with the highest bid wins, but ties are broken asymmetrically among the players (i.e., some players have a pre-defined priority over the others when they send the same message). We show how to derive the optimal values for the parameters of the priority game. These optimal mechanisms are asymmetric by definition, although the players are a priori identical. The asymmetry is in contrast with optimal mechanisms with unconstrained communication, where symmetric mechanisms achieve optimal welfare (the second-price auction, Vickrey, 1961) and optimal profit (the Myerson auction, Myerson, 1981, for symmetric bidders). Furthermore, we show that for any number of players, as the allowed number of messages grows, the loss due to bounded communication is in order of $O(\frac{1}{k^2})$. The bound is tight for some distributions of valuations (e.g., for the uniform distribution). In addition, we consider the case in which the number of players grows while each player has exactly two possible messages. We show that priority games are optimal for this case as well, and we also characterize the parameters for the optimal mechanisms and show that they can be generated from a simple recursive formula. We offer an asymptotic bound on the welfare and profit losses due to bounded communication as the number of players grows (it is $O(\frac{1}{n})$ for the uniform distribution). All the optimal mechanisms in this paper are deterministic, but they are optimal even if the auctioneer is allowed to randomize.[2]

---

1. There have been several other studies considering various computational considerations in auction design: timing (e.g., Lavi & Nisan, 2004; Roth & Ockenfels, 2002), unbounded supply (e.g., Feigenbaum, Papadimitriou, & Shenker, 2001; Goldberg, Hartline, & Wright, 2001; Bar-Yossef, Hildrum, & Wu, 2002), computational complexity in combinatorial auctions (see the survey by Cramton, Shoham, & Steinberg, 2006) and more.

Our analysis implies some expected as well as some unexpected results:

- **Low welfare and profit loss:** Even severe bounds on communication result in only a mild loss of efficiency. We present mechanisms in which the welfare loss and the profit loss decrease exponentially in the number of the communication bits (and quadratically in the number $k$ of the allowed bids). For example, with two bidders whose valuations are uniformly distributed on $[0, 1]$, the optimal 1-bit auction brings expected welfare 0.648, compared to the first-best expected welfare 0.667.

- **Asymmetry helps:** Asymmetric auctions are better than symmetric ones with the same communication bounds. For example, with two bidders whose valuations are uniformly distributed in $[0, 1]$, symmetric 1-bit auctions only achieve expected welfare of 0.625, compared to 0.648 for asymmetric ones. We prove that both welfare- and profit-maximizing auctions must be discriminatory in both allocation and payments.

- **Dominant-strategy incentive compatibility is achieved at no additional cost:** The auctions we design have dominant-strategy equilibria and are ex-post individually rational[3], yet are optimal even without any incentive constraints (for welfare maximization), or among all Bayesian-Nash incentive-compatible and interim individually rational auctions (for profit maximization). This generalizes well-known results for the case without any communication constraints.

- **Bidding using "mutually-centered" thresholds is optimal:** We show that in the optimal auctions with $k$ messages, bidders simply partition the range of valuations into $k$ interval ranges and announce their interval. In 2-bidder mechanisms, each threshold will have the interesting property of being the average value of the other bidder in the respective interval. We denote such threshold vectors as "mutually centered".

- **Sequential mechanisms can do better, but only up to a linear factor:** Allowing players to send messages sequentially rather than simultaneously can achieve a higher payoff than in simultaneous mechanisms. However, the payoff in any such multi-round mechanism among $n$ players can be achieved by a simultaneous mechanism in which the players send messages which are longer only by a factor of $n$. This result is surprising in light of the fact that in general the restriction to simultaneous communication can increase communication complexity exponentially. Our results for sequential mechanisms are very robust in several aspects. They allow the players to send messages of various sizes and in any order, and they allow the auctioneer to adaptively determine the order and the size of the messages based on the history of the messages. The auctioneer may also use randomized decisions.

---

2. The mechanisms are optimal even when the auctions are run repeatedly, as long as the bidders' values are uncorrelated over time.

3. A mechanism is ex-post individually rational if a player never pays more than her value. Interim individual rationality is a weaker property, in which a player will not pay more than his value *on average*. Individual rationality constraints are essential for the study of revenue maximization (otherwise, the potential revenue is unbounded).

Although the welfare-maximizing mechanisms are asymmetric, symmetric mechanisms can also be close to optimal: we show that as the number $k$ of possible messages grows, while the number of players is fixed, the loss in optimal symmetric mechanisms converges to zero at the same rate as the loss in efficient "priority games". The optimal loss in symmetric and asymmetric mechanisms, however, differs by a constant factor. On the other hand, when we fix the number of messages, we show that the optimal loss in asymmetric mechanisms converges to zero *asymptotically faster* than in optimal symmetric mechanisms ($O(\frac{1}{n})$ compared to $O(\frac{\log n}{n})$, for the uniform distribution).

We now demonstrate the properties above with an example for the simplest case: a 2-bidder mechanism where each player has two possible bids (i.e., 1 bit) and the values are distributed uniformly.

**Example 1.** *Consider two players, Alice and Bob, with values uniformly distributed between* $[0, 1]$. *A 1-bit auction among these players can be described by a 2x2 matrix, where Alice chooses a row, and Bob chooses a column. Each entry of the matrix specifies the allocation and payments given a combination of bids. The mechanism is allowed to toss coins to determine the allocations. Figure 1 describes an example for such a mechanism, and we denote this mechanism by* $g_1$.

*A* strategy *defines how a player determines his bid according to his private value. We first note that in* $g_1$, *both players have dominant strategies, i.e., strategies that are optimal regardless of the actions of the other players. Consider the following threshold strategy: "bid 1 if your valuation is greater than* $\frac{1}{3}$, *else bid 0". Clearly, this strategy is dominant for Alice in* $g_1$: *when her valuation is smaller than* $\frac{1}{3}$ *she will gain a negative utility if she bids "1"; When her valuation is greater than* $\frac{1}{3}$, *bidding "0" gives her a utility of zero, but she can get positive utility by bidding "1". Similarly, a threshold strategy with the threshold* $\frac{2}{3}$ *is dominant for Bob.*

*The* social welfare *in a mechanism measures the total happiness of the players from the outcome, or in our case, the value of the player that receives the item. The expected welfare in* $g_1$, *given that the players follow their dominant strategies, is easily calculated to be* $\frac{35}{54} = 0.648$: *both players will bid "0" with probability* $\frac{1}{3} \cdot \frac{2}{3}$, *and the expected welfare in this case equals the expected value of Bob,* $\frac{1}{2} \cdot \frac{2}{3}$. *Similar computations show that the expected welfare is indeed:*

$$\frac{1}{3}\frac{2}{3}\frac{\left(\frac{2}{3}\right)}{2} + \frac{1}{3}(1 - \frac{2}{3})\frac{\left(1 + \frac{2}{3}\right)}{2} + (1 - \frac{1}{3})\frac{2}{3}\frac{\left(1 + \frac{1}{3}\right)}{2} + (1 - \frac{1}{3})(1 - \frac{2}{3})\frac{\left(1 + \frac{2}{3}\right)}{2} = \frac{35}{54}$$

*We see that despite restricting the communication from an infinite number of bits to a single bit only, a relatively small welfare loss of* $\frac{1}{54}$ *was incurred. Of course, a random allocation that can be implemented without communication at all will result in an expected welfare of* $\frac{1}{2}$, *and this may be regarded as our naive benchmark.*

*It turns out that the mechanism described in Figure 1 maximizes the expected welfare: no other 1-bit mechanism achieves strictly higher expected welfare with any pair of bidders' strategies (that is, regardless of the concept of equilibrium we use). We note that the optimal mechanism is asymmetric (a "priority game") – ties are always broken in favor of Bob, and that this mechanism is optimal even when randomized decisions are allowed. Note that the optimal symmetric 1-bit mechanism uses randomization, but only achieves an expected welfare of 0.625 (the mechanism is illustrated in Appendix A.3 and see also Footnote 20).*

*Finally, we note that the optimal thresholds of the players are "mutually centered". That is, Alice's value $\frac{1}{3}$ is the average value of Bob when he bids 0 and Bob's value $\frac{2}{3}$ is the average value of Alice when she bids 1. The intuition is simple: given that Bob bids "0", his average value is $\frac{1}{2} \cdot \frac{2}{3} = \frac{1}{3}$. For which values of Alice should an efficient mechanism give her the item? Clearly when her value is greater than the average value of Bob. Therefore, Alice should use the threshold $\frac{1}{3}$.*

The most closely related work in the economic literature is by Harstad and Rothkopf (1994), who considered similar questions in cases of restricting the bid levels in oral auctions to discrete levels, and by Wilson (1989) and McAfee (2002) who analyzed the inefficiency caused by discrete priority classes of buyers. In particular, Wilson showed that as the number $k$ of priority classes grows, the efficiency loss is asymptotically proportional to $\frac{1}{k^2}$. While in the work of Wilson the buyers' aggregate demand is known while supply is uncertain, in our model the demand is uncertain. Both Wilson and Harstad and Rothkopf restrict attention to symmetric mechanisms, while we show that creating endogenous asymmetry among ex ante identical buyers is beneficial. Another related work is by Bergemann and Pesendorfer (2001), where the seller can decide on the accuracy by which bidders know their private values. This problem is different than ours, since the bidders in our model know their valuations. The work by Parkes (2005) is also related. He compared the efficiency of simultaneous and sequential auctions under uncertainties on the values of the players. Recent work also studied similar discrete-bid model in the context of ascending auctions and auctions that use take-it-or-leave-it offers (Kress & Boutilier, 2004; Sandholm & Gilpin, 2006; David, Rogers, Schiff, Kraus, & Jennings, 2005).

The organization of the paper is as follows: Section 2 presents our model definition and introduces our notations and Section 3 presents a characterization of the welfare- and profit-optimal 2-bidder auctions. Section 4 characterizes optimal mechanisms with an arbitrary number of bidders, but 2 possible bids for each player. In Section 5 we give an asymptotic analysis of the minimal welfare and profit losses in the optimal mechanisms. Finally, Section 6 compares simultaneous and sequential mechanisms with bounded communication.

## 2. The Model

This section presents our formal model and the notations we use.

| $\begin{matrix} & B \\ A & \end{matrix}$ | 0 | 1 |
|---|---|---|
| 0 | $B$ wins and pays 0 | $B$ wins and pays 0 |
| 1 | $A$ wins and pays $\frac{1}{3}$ | $B$ wins and pays $\frac{2}{3}$ |

Figure 1: $(g_1)$ A 2-bidder 1-bit game that achieves maximal expected welfare. For example, when Alice (the rows bidder) bids "1" and Bob bids "0", Alice wins and pays $\frac{1}{3}$.

## 2.1 The Bidders and the Mechanism

We consider single item, sealed bid auctions among $n$ risk-neutral players. Player $i$ has a private valuation for the object $v_i \in [\underline{a}, \overline{b}]$.[4] The valuations are independently drawn from cumulative probability functions $F_i$. In some parts of our analysis[5], we assume the existence of an always-positive probability density function $f_i$. We will sometime treat the seller as one of the bidders, numbered 0. The seller has a constant valuation $v_0$ for the item. We consider a normalized model, i.e., bidders' valuations for not having the item are $\underline{a}$.

The novelty in our model, compared to the standard mechanism-design settings, is that each bidder $i$ can send a message of $t_i = \lg(k_i)$ **bits** to the mechanism, i.e., player $i$ can choose one of possible $k_i$ **bids** (or messages). Denote the possible set of bids for bidder $i$ as $\beta_i = \{0, 1, 2, ..., k_i - 1\}$. In each auction, bidder $i$ chooses a bid $b_i \in \beta_i$. A mechanism should determine the allocation and payments given a vector of bids $b = (b_1, ..., b_n)$:

**Definition 1.** *A mechanism $g$ is composed of a pair of functions $(a, p)$ where:*

- *$a : (\beta_1 \times ... \times \beta_n) \to [0, 1]^{n+1}$ is the allocation scheme (not necessarily deterministic). We denote the $i$th coordinate of $a(b)$ by $a_i(b)$, which is bidder $i$'s probability for winning the item when the bidders bid $b$. Clearly, $\forall i \; \forall b \; a_i(b) \geq 0$ and $\forall b \; \sum_{i=0}^{n} a_i(b) = 1$. If $a_0(b) > 0$, the seller will keep the item with a positive probability.*

- *$p : (\beta_1 \times ... \times \beta_n) \to \Re^n$ is the payment scheme. $p_i(b)$ is the payment of the $i$th bidder given a vector of bids $b$.[6]*

**Definition 2.** *In a mechanism with $k$ possible bids, for every bidder $i$, $|\beta_i| = k_i = k$. We denote the set of all the mechanisms with $k$ possible bids among $n$ bidders by $G_{n,k}$. We denote the set of all the $n$-bidder mechanisms in which $|\beta_i| = k_i$ for each bidder $i$, by $G_{n,(k_1,...,k_n)}$.*

A *strategy* $s_i$ for bidder $i$ in a game $g \in G_{n,(k_1,...,k_n)}$ describes how a bidder determines his bid according to his valuation, i.e., it is a function $s_i : [\underline{a}, \overline{b}] \to \{0, 1, ..., k_i - 1\}$. Let $s_{-i}$ denote the strategies of the bidders except $i$, i.e., $s_{-i} = (s_1, ..., s_{i-1}, s_{i+1}, ..., s_n)$. We sometimes use the notation $s = (s_i, s_{-i})$.

---

4. For simplicity, we use the range $[0, 1]$ in some parts of the paper. Using the general interval will be required, though, for the characterization of the optimal mechanisms, mainly due to the reduction we use for maximizing the revenue that translates the original support to their virtual valuations that are drawn from another interval.

5. That is, in the characterization of the optimal mechanisms in Sections 3.2 and 4 and when using the concept of *virtual valuation* in Sections 3.3 and 5.2

6. Note that we allow non-deterministic allocations, but we ignore non-deterministic payments (since we are interested in expected values, using lottery for the payments has no effect on our results).

**Definition 3.** *A real vector* $(t_0, t_1, ..., t_k)$ *is a vector of threshold values if* $t_0 \leq t_1 \leq ... \leq t_k$.

**Definition 4.** *A strategy* $s_i$ *is a threshold strategy based on a vector of threshold values* $(t_0, t_1, ..., t_k)$, *if for every bid* $j \in \{0, ..., k_i - 1\}$ *and for every valuation* $v_i \in [t_j, t_{j+1})$, *bidder* $i$ *bids* $j$ *when his valuation is* $v_i$, *i.e.,* $s_i(v_i) = j$ *(and for every* $v_i, v_i \in [t_0, t_k]$). *We say that* $s_i$ *is a threshold strategy, if there exists a vector* $t$ *of threshold values such that* $s_i$ *is a threshold strategy based on* $t$.

## 2.2 Optimality Criteria

The bidders aim to maximize their (quasi-linear) utilities. The utility of bidder $i$ is $\underline{a}$ when he loses (and pay nothing), and $v_i - p_i$ when he wins and pay $p_i$. Let $u_i(g, s)$ denote the expected utility of bidder $i$ from a game $g$ when the bidders use the vector of strategies $s$ (implicit here is that this utility depends on the value $v_i$).

**Definition 5.** *A strategy* $s_i$ *for bidder* $i$ *is dominant in a mechanism* $g \in G_{n,(k_1,...,k_n)}$ *if regardless of the other bidders' strategies* $s_{-i}$, *$i$ cannot increase his expected utility by a deviation to another strategy, i.e.,*

$$\forall \widetilde{s_i} \ \forall s_{-i} \ \ u_i(g, (s_i, s_{-i})) \geq u_i(g, (\widetilde{s_i}, s_{-i}))$$

**Definition 6.** *A profile of strategies* $s = (s_1, ..., s_n)$ *forms a Bayesian-Nash equilibrium (BNE) in a mechanism* $g \in G_{n,(k_1,...,k_n)}$, *if for every bidder* $i$, *$s_i$ is the best response for the strategies* $s_{-i}$ *of the other bidders, i.e.,*

$$\forall i \ \forall \widetilde{s_i} \ \ u_i(g, (s_i, s_{-i})) \geq u_i(g, (\widetilde{s_i}, s_{-i}))$$

We use standard participation constraints definitions: We say that a profile of strategies $s = (s_1, ..., s_n)$ is *ex-post individually rational* in a mechanism $g$, if every bidder never pays more than his actual valuation (for any realization of the valuations); we will assume a strong version of this definition that holds even in randomized mechanisms. We say that a strategies profile $s = (s_1, ..., s_n)$ is *interim individually rational* in a mechanism $g$ if every bidder $i$ achieves a non-negative *expected* utility, given any valuation he might have, when the other bidders play with $s_{-i}$.

Our goal is to find optimal, communication-bounded mechanisms. As the mechanism designers, we will try to optimize "social" criteria such as *welfare* (efficiency) and the seller's *profit*.

The *expected welfare* from a mechanism $g$, when bidders use the strategies $s$, is the expected social surplus. Because the item is indivisible, the social surplus is actually the valuation of the bidder who receives the item. If the seller keeps the item, the social welfare is $v_0$.

**Definition 7.** *Let* $w(g, s)$ *denote the expected welfare (or expected efficiency) in the $n$-bidder game $g$ when the bidders' strategies are $s$, i.e., the expected value of the player (possibly the seller) who receives the item in $g$. Let* $w^{opt}_{n,(k_1,...,k_n)}$ *denote the maximal possible expected* *welfare from any n-bidder game where each bidder $i$ has $k_i$ possible bids, with any vector of strategies allowed, i.e.,*

$$w_{n,(k_1,\ldots,k_n)}^{opt} = \max_{g \in G_{n,(k_1,\ldots,k_n)}, \ s} w(g,s)$$

*When all bidders have $k$ possible bids we use the notation $w_{n,k}^{opt} = w_{n,(k,\ldots,k)}^{opt}$*

Actually, the optimal welfare should have been defined as the maximum expected welfare that can be obtained *in equilibrium*. Since we later show that the optimal welfare without strategic considerations is dominant-strategy implementable, we use the above definition for simplicity. Note that even in the absence of communication restrictions, optimizing the welfare objective is obtained by a first-best solution (using the VCG scheme); profit maximization, on the other hand, is obtained by a second-best solution (incentive constraints bind in Myerson's auction, Myerson, 1981).

**Definition 8.** *The seller's profit is the payment received from the winning bidder, or $v_0$ when the seller keeps the item.[7] Let $r(g,s)$ denote the expected profit in the n-bidder game $g$ where the bidders' strategies are $s$. Let $r_{n,k}^{opt}$ denote the maximal expected profit from an n-bidder mechanism with $k$ possible bids and some vector of interim individually-rational strategies $s$ that forms a Bayesian-Nash equilibrium in $g$:*

$$r_{n,k}^{opt} = \max_{\substack{g \in G_{n,k} \\ s \text{ is interim IR and in BNE in } g}} r(g,s)$$

Note that we define the optimal welfare as the maximal welfare among all mechanisms and strategies, not necessarily in equilibria, and we define the optimal profit as the maximal profit achievable in interim-IR Bayesian-Nash equilibria in any mechanism. Yet, the optimal mechanisms (for both measures) that we present in this paper implement these optimal values with dominant strategies and ex-post IR.[8]

**Definition 9.** *We say that a mechanism $g \in G_{n,k}$ achieves the optimal welfare (resp. profit), if $g$ has an interim-IR Bayesian-Nash equilibrium $s$ for which the expected welfare (resp. profit) is $w(g,s) = w_{n,k}^{opt}$ (resp. $r(g,s) = r_{n,k}^{opt}$ ).*

*We say that a mechanism $g \in G_{n,k}$ incurs a welfare loss (resp. profit loss) of $L$, if it achieves an expected welfare (resp. profit) which is additively smaller than the optimal welfare (resp. profit) with unbounded communication by $L$ (the optimal results with unbounded communications are the best results achievable with interim-IR Bayesian-Nash equilibria).*

## 3. Optimal Mechanisms for Two Bidders

In this section we present 2-bidder mechanisms with bounded communication that achieve optimal welfare and profit. In Section 4 we will present the characterization of the welfare-optimal and profit-optimal $n$-bidder mechanisms with 2 possible bids for each bidder. The characterization of the optimal mechanisms in the most general case ($n$ bidders and $k$ possible bids) remains an open question. Anyway, our asymptotic analysis of the optimal welfare loss and the profit loss (in Section 5) holds for the general case, and shows *asymptotically optimal* mechanisms.

---

7. When $v_0 = 0$, the expected profit is equivalent to the seller's expected *revenue*.

8. Note that ex-ante IR, i.e., when bidders do not know their type when choosing their strategies, is non-interesting in this model, since the auctioneer can then simply ask each bidder to pay her expected valuation.

We first show that the allocation rules in efficient mechanisms have a certain structure we call *priority games*. The term priority game means that the allocation rule uses an asymmetric tie breaking rule: the winning player is the player with the highest priority among the bidders that bid the highest. One consequence is that the bidder with the lowest priority will win only when his bid is strictly higher than all other bids. Note that the term "priority game" refers to the asymmetry in the mechanism's allocation function, but additional asymmetry will also appear in the payment scheme. A *modified priority game* has a similar allocation, except the item is not allocated when all the bidders bid their lowest bid.[9] We will mostly be interested in such mechanisms when the players have the same bid space $\beta_i$.

**Definition 10.** *A game is called a priority game if it allocates the item to the bidder $i$ that bids the highest bid (i.e., when $b_i > b_j$ for all $j \neq i$, the allocation is $a_i(b) = 1$ and $a_j(b) = 0$ for $j \neq i$), with ties consistently broken according to a pre-defined order on the bidders).*

*A game is called a modified priority game if it has an allocation as of a priority game, except when all bidders bid 0, the seller keeps the item.*

It turns out to be useful to build the payment scheme of such mechanisms according to a given profile of threshold strategies:

**Definition 11.** *An n-bidder priority game based on a profile of threshold values' vectors $\overrightarrow{t} = (t^1, ..., t^n) \in \times_{i=1}^n \Re^{k+1}$ (where for every $i$, $t_0^i \leq t_1^i \leq ... \leq t_k^i$) is a mechanism whose allocation is a priority game and its payment scheme is as follows: when bidder $j$ wins the item for a vector of bids $b$ she pays the smallest valuation she might have and still win the item, given that she uses the threshold strategy $s_j$ based on $t^j$, i.e, $p_j(b) = \min\{v_j | a_j(s_j(v_j), b_{-j}) = 1\}$. We denote this mechanism as $PG_k(\overrightarrow{t})$. A modified priority game with a similar payment rule is called a modified priority game based on a profile of threshold-value vectors, and is denoted by $MPG_k(\overrightarrow{t})$.*

For 2-bidder games, we may use the notations $PG_k(x, y)$, $MPG_k(x, y)$ (where $x, y$ are some vectors of threshold values). The mechanisms $PG_k(x, y)$ and $MPG_k(x, y)$ are presented in Figure 2. Note $PG_k(x, y)$ and $MPG_k(x, y)$ differ only when bidder $A$ bids "0" (i.e., the first line of the game's matrix).

We now observe that priority games and the modified priority games, with the payments schemes that were described above, have two desirable properties: they admit a dominant-strategy equilibrium, and they are ex-post individually rational when the players follow these dominant strategies.

As for the dominant strategies, a well known result in mechanism design (see Mookherjee & Reichelstein, 1992 and also Lemma 1 in Segal, 2003) states that for any monotone[10] allocation rule there is *some* transfer (i.e., payment) rule that would implement the desired allocation in dominant strategies. For deterministic auctions, to support this equilibrium, each winning bidder should pay the smallest valuation for which she still wins (fixing the behavior of the other bidders). The payments in Definition 11 are defined in this way, and therefore they support the dominant-strategy implementation. It follows that the threshold strategies based on the threshold values vector $\overrightarrow{t}$ are dominant in both $PG_k(\overrightarrow{t})$ and $MPG_k(\overrightarrow{t})$. It is clear from the definition of priority games and modified priority games that, when playing their dominant threshold strategies, winning players will never pay more than their value, and losing players will pay zero. Ex-post IR follows.

---

9. Modified priority games can be viewed as priority games that treat the seller as one of the bidders with the lowest "priority" (then, the seller always bids his second-lowest bid).

10. A mechanism is monotone if the probability that some bidder wins increases as he raises his bid, fixing the bids of the other bidders. See Definition 12 below for our model.

Actually, the observation about the payments that lead to dominant strategies is even more general. We observe that monotone mechanisms reveal enough information, despite the communication constraints, to find transfer rules that support the dominant-strategy implementation. Therefore, when characterizing the optimal mechanisms we can focus on defining monotone allocation schemes under the communication restrictions, and the transfers that lead to dominant-strategy equilibria can be concluded "for free". In other words, we can use the 2-stage approach that is widely used in the mechanism-design literature also for bounded-communication settings: first solve the optimal allocation rule, and then construct transfers that satisfy the desired incentive-compatibility and individual-rationality constraints.

*Remark* 1. This argument holds for more general environments: in environments in which each player has a one-dimensional private value and a quasi-linear utility, if a non-monetary allocation rule can be implemented in dominant strategies with some transfers, then any communication protocol[11] realizing this rule also reveals enough information to construct supporting transfers for the dominant strategies. To see this, recall that in direct-revelation mechanisms (i.e., with unbounded communication), if the allocation rule proves to be monotonic, there are transfers that support a dominant-strategy equilibrium. The transfers will be defined according to some allocation-dependent thresholds, e.g., for a deterministic allocation rule every bidder should pay the smallest valuation for which she still wins. By standard revelation-principle arguments, any monotonic allocation rule in *bounded communication* mechanisms, can be viewed as a monotonic direct-revelation mechanism with unbounded communication, and therefore such supporting transfers exist. The supporting transfers are determined by the changes in the allocation rule as the valuation of each bidder increases, so the transfers change as the allocation rule changes. Thus, with the same communication protocol that is used for determining the allocation, we can reveal the transfers that support a dominant-strategy implementation.

## 3.1 The Efficiency of Priority Games

The characterization of the welfare-maximizing mechanism is done in two steps: we first show that the allocation scheme in 2-bidder priority games is optimal[12]. Afterwards, we will characterize the strategies of the players that lead to welfare maximization in priority games; this will complete the description of the outcome of the mechanism for every profile of bidder valuations. These two stages do not take strategic behavior of the bidders into account. Yet, as observed before, since the allocation scheme is proved to be monotone, there exists a payment scheme for which these strategies are dominant.

---

11. Here we deal with simultaneous communication, i.e., where all bidders send their messages simultaneously. Our observation is not true for sequential mechanisms (see Section 6).
12. We assume, w.l.o.g., throughout this paper that in 2-bidder priority games $B \succ A$, i.e., the mechanism allocates the item to $A$ if she bids higher than $B$, and otherwise to $B$.

|     | 0        | 1        | ... | k-2          | k-1          |
| --- | -------- | -------- | --- | ------------ | ------------ |
| 0   | $B, y_0$ | $B, y_0$ | ... | $B, y_0$     | $B, y_0$     |
| 1   | $A, x_1$ | $B, y_1$ | ... | $B, y_1$     | $B, y_1$     |
| 2   | $A, x_1$ | $A, x_2$ | ... | $B, y_2$     | $B, y_2$     |
| ... | ...      | ...      | ... | ...          |              |
| k-2 | $A, x_1$ | $A, x_2$ | ... | $B, y_{k-2}$ | $B, y_{k-2}$ |
| k-1 | $A, x_1$ | $A, x_2$ | ... | $A, x_{k-1}$ | $B, y_{k-1}$ |

|     | 0        | 1        | ... | k-2          | k-1          |
| --- | -------- | -------- | --- | ------------ | ------------ |
| 0   | $\phi$   | $B, y_1$ | ... | $B, y_1$     | $B, y_1$     |
| 1   | $A, x_1$ | $B, y_1$ | ... | $B, y_1$     | $B, y_1$     |
| 2   | $A, x_1$ | $A, x_2$ | ... | $B, y_2$     | $B, y_2$     |
| ... | ...      | ...      | ... | ...          |              |
| k-2 | $A, x_1$ | $A, x_2$ | ... | $B, y_{k-2}$ | $B, y_{k-2}$ |
| k-1 | $A, x_1$ | $A, x_2$ | ... | $A, x_{k-1}$ | $B, y_{k-1}$ |

Figure 2: A priority game (left) and a modified priority game (right) both based on the threshold values' vectors $x$, $y$. In each entry, the left argument denotes the winning bidder, and the right argument is the price she pays. The mechanisms differ in the allocation for all-zero bids, and the payments in the first row.

**Definition 12.** *A mechanism $g \in G_{n,k}$ is monotone if for any vector of bids $b$ and for any bidder $i$, the probability that bidder $i$ wins the item cannot decrease when only his bid increases, i.e.,*

$$\forall b \quad \forall i \quad \forall b_i^{'} > b_i \quad a_i(b_i, b_{-i}) \leq a_i(b_i^{'}, b_{-i})$$

In the following theorem we prove that priority games are welfare maximizing. The proof is composed of four steps: We first show that we can assume that the bidders in the optimal mechanisms use threshold strategies. Then, we show that the allocation in the optimal mechanisms is, w.l.o.g., monotone and deterministic. We then show that the optimal mechanisms do not "waste" communication, i.e., no two "rows" or two "columns" in the allocation matrix of the optimal mechanism are identical. Finally, we use these properties, together with several combinatorial arguments, to derive the optimality of priority games.

**Theorem 3.1.** *(Priority games' efficiency) For every pair of distribution functions of the bidders' valuations, and for every $v_0$, the optimal welfare (i.e., $w_{2,k}^{opt}$ ) is achieved in either a priority game or a modified priority game (with some pair of threshold strategies).*

*Proof.* We first prove the theorem given that the seller has a low reservation value, i.e., $v_0 \leq \underline{a}$. Recall that at this point we aim to find the welfare-maximizing allocation scheme, without taking the incentives of the bidders into account. The proof uses the following three claims. For a later use, Claims 3.2 and 3.3 are proved for $n$ players.

**Claim 3.2.** *(Optimality of threshold strategies) Given any mechanism $g \in G_{n,(k_1,...,k_n)}$, there exists a vector of threshold strategies $s$ that achieve the optimal welfare in $g$ among all possible strategies, i.e., $w(g, s) = \max_{\widetilde{s}} w(g, \widetilde{s})$.*

*Proof.* (sketch - a formal proof is given in Appendix A.1)

Given a profile of welfare-maximizing strategies in $g$, we can modify the strategy of each bidder (w.l.o.g., bidder 1) to be a threshold strategy maintaining at least the same expected welfare. The idea is that fixing the strategies $s_{-1}$ of the other bidders, the expected welfare achieved when bidder 1 bids some bid $b_1$ is a linear function in bidder $i$'s value $v_1$. The maximum of all these linear functions is a piecewise-linear function, and it specifies the optimal welfare as a function of $v_1$. Bidder 1 can use a threshold strategy according to the breaking points of this piecewise-linear function that choose the welfare-maximizing linear function at each segment. Clearly, there are at most $k - 1$ breaking points. □

**Claim 3.3. (Optimality of deterministic, monotone mechanisms)** *For every $n$ and $k_1, ..., k_n$, there exists a mechanism $g \in G_{n,(k_1,...,k_n)}$ with optimal welfare (i.e., there exists a profile $s$ of strategies such that $w(g, s) = w^{opt}_{n,(k_1,...,k_n)}$) which is monotone, deterministic (i.e., the winner is fixed for each combination of bids) and in which the seller never keeps the item.*

*Proof.* Consider a mechanism $g \in G_{n,(k_1,...,k_n)}$ and a profile $s$ of strategies that maximize the expected welfare, that is, $w(g, s) = w^{opt}_{n,(k_1,...,k_n)}$. A social planner, aiming to maximize the welfare, will always allocate the item to the bidder with the highest expected valuation. That is, for each combination of bids $b = (b_1, .., b_n)$ we will allocate the item (i.e., $a_i(b) = 1$) to a bidder $i$ such that $i \in argmax_j(E(v_j|s_j(v_j) = b_j))$. The expected welfare clearly did not decrease. In addition, we always allocate the item (we assume that $v_0 \leq \underline{a}$), and the allocation is deterministic. Finally, we can assume, w.l.o.g., that for each bidder $i$ the bids' names (i.e., "0","1" etc.) are ordered according to the expected value this bidder has. Then, the mechanism will also be monotone: if a winning bidder $i$ increases his bid, his expected valuation will also increase, while the expected welfare of all the other bidders will not change. Thus, bidder $i$ will still have the maximal expected valuation. □

**Claim 3.4. (Additional bids strictly help)** *Consider a deterministic, monotone mechanism $g \in G_{2,k}$ in which the seller never keeps the item. If $g$ achieves the optimal expected welfare, then in the matrix representation of $g$ no two rows (or columns) have an identical allocation scheme.*

*Proof.* The idea that an optimal protocol exploits all its communication resources is intuitive, although it does not hold in all settings (a trivial example is calculating the parity of two binary numbers, more involved examples can be found in Kushilevitz & Nisan, 1997). We do not have a simple proof for this statement in our model, and the proof is based on Lemma A.1 in the appendix in the following way: Consider such an optimal mechanism $g \in G_{2,k}$ with two identical rows. This mechanism achieves the optimal welfare when the players use some profile of strategies $s$. $g$'s monotonicity implies that the two identical rows are adjacent. Thus, there is a mechanism with $\widetilde{g} \in G_{2,(k-1,k)}$ with $k - 1$ possible bids for the rows bidder that achieves exactly the same expected welfare as $g$ (when the identical rows are united to one). This welfare is achieved with the same strategies $s$ of the bidders, where the rows player bids the united row instead of the two identical rows. The claim will now follow from Lemma A.1 in the appendix. According to this lemma, the optimal welfare from a game where both bidders have $k$ possible bids cannot be achieved when one of the bidders has only $k - 1$ possible bids (i.e., $w^{opt}_{2,k} > w^{opt}_{2,(k-1,k)}$). □

Now, due to Claim 3.3, there is a deterministic, monotone game in which the item must be sold that achieves $w^{opt}_{2,k}$. In such games, the allocation scheme in some row $i$ looks like $[A, ..., A, B...B]$. Due to Claim 3.4, in the matrix representation of this optimal game, there are no two rows with the same allocation scheme. There are $k+1$ possible monotone rows for the game matrix (with prefix of 0 to $k$ A's), but our mechanism has only $k$ rows. Similarly, we have $k$ different columns (of possible $k+1$) in the mechanism. Assume that the row $[B, B, ..., B]$ is in $g$. Then, the column $[A, A, ..., A]$ is clearly not in $g$. Therefore, our game matrix consists of all the columns except $[A, A, ..., A]$, which compose the priority game where $B \succ A$. If the row $[B, B, ..., B]$ is not in $g$, then $g$ is the priority game where $A \succ B$.

Next, we complete the proof for any seller's valuation $v_0$. Consider a mechanism $h \in G_{2,k}$ and a pair of threshold strategies based on some threshold-value vectors $\widetilde{x}, \widetilde{y}$ that achieve the optimal welfare among all mechanisms and strategies (due to Claim 3.2, such strategies exist). We will modify $h$, such that the expected welfare (with $\widetilde{x}, \widetilde{y}$) will not decrease. Let $a$ be the smallest index such that $E(v_A | \widetilde{x_a} \le v_A \le \widetilde{x_{a+1}}) \ge v_0$. Let $b$ be the smallest index such that $E(v_B | \widetilde{y_b} \le v_B \le \widetilde{y_{b+1}}) \ge v_0$. If $a = 0$ or $b = 0$, the item is never allocated to the seller, and the efficient mechanism is as if $v_0 \le \underline{a}$.

When $a, b > 0$, consider some vector of bids $(i, j)$. When $i < a$ and $j < b$, the expected valuations of both $A$ and $B$ are smaller than $v_0$. Thus, the seller should keep the item for optimal welfare. When $i < a$ and $j \ge b$, the expected welfare of bidder $B$ is above $v_0$, and $A$'s expected welfare among all mechanisms is below $v_0$, thus we can allocate the item to $B$ and the welfare will not decrease. Similarly, we should allocate the item to $A$ when $i \ge a$ and $j < b$. When $i < a$, the allocation is done regardless to $i$, thus we can assume that $x_a$ is the first threshold (i.e., $a = 1$), and similarly $b = 1$.

Now, we show the optimal allocation for combinations of bids $(i, j)$ such that $i \ge a$ and $j \ge b$. Here, the item will not be allocated to the seller, so we actually perform an auction with $k - 1$ possible bids for each bidder, when the bidders' valuation are in the range $[\widetilde{x_1}, 1]$, $[\widetilde{y_1}, 1]$. Note that the proof (above) for the case of $v_0 \le \underline{a}$ holds for such ranges, so the optimal welfare is achieved in a priority game. Altogether, the optimal mechanism turns out to be a modified priority game. □

## 3.2 Efficient 2-bidder Mechanisms with $k$ Possible Bids

Now, we can finally characterize the efficient mechanisms in our model. It turns out that the optimal threshold values for priority games are *mutually centered*, i.e., each threshold is the expected valuation of *the other* bidder, given that the valuation of the other bidder lies between his two adjacent thresholds.

**Definition 13.** *The threshold values* $x = (x_0, x_1, ..., x_{k-1}, x_k)$, $\quad y = (y_0, y_1, ..., y_{k-1}, y_k)$ *for bidders* $A, B$ *respectively are mutually centered, if the following constraints hold:*

$$\forall 1 \le i \le k-1 \quad x_i = E(v_B \,|\, y_{i-1} \le v_B \le y_i) = \frac{\int_{y_{i-1}}^{y_i} f_B(v_B) \cdot v_B \, dv_B}{F_B(y_i) - F_B(y_{i-1})}$$

$$\forall 1 \le i \le k-1 \quad y_i = E(v_A \,|\, x_i \le v_A \le x_{i+1}) = \frac{\int_{x_i}^{x_{i+1}} f_A(v_A) \cdot v_A \, dv_A}{F_A(x_{i+1}) - F_A(x_i)}$$

It is easy to see that given any pair of distribution functions, a pair $\overrightarrow{x}, \overrightarrow{y}$ of mutually-centered vectors is uniquely defined (when $x_k = y_k$ and, w.l.o.g., $y_1 \ge x_1$). The basic idea is that if $x_1$ is known, we can clearly calculate $y_1$ (the smallest value that solves $x_1 = E_{v_B}(v_B | y_0 \leq v_B \leq y_1)$). Similarly, it is easy to see that all the variables $x_i$ and $y_i$ can be considered as continuous, monotone functions of $x_1$. Now, let $z$ be the solution for the equation $y_{k-1} = E(v_A | x_{k-1} \leq v_A \leq z)$. For satisfying all the $2(k-1)$ equations, $z$ must equal $x_k$. Since $z$ is also a continuous monotone function of $x_1$, there is only a single value of $x_1$ for which all the equations hold.

The following intuition shows why the optimal thresholds in priority games must be mutually centered: Assume that Alice bids $i$, that is, her value is in the range $[x_i, x_{i+1}]$. In a monotone mechanism, the mechanism designer has to decide what is the minimal value for which Bob wins when Alice bids $i$. If the value of Bob is at least the average value of Alice, given that she bids $i$, then Bob should clearly receive the item. Therefore, Bob's threshold will be exactly this expected value of Alice. The proof has to handle few subtleties for which the intuition above does not suffice (like the characterization of the first thresholds in the optimal modified priority games, see below), thus we will derive the mutually-centered condition from the solution of the optimization problem.

Let $x^w = (\underline{a} = x_0^w, x_1^w, ..., x_{k-1}^w, x_k^w = \overline{b})$ and $y^w = (\underline{a} = y_0^w, y_1^w, ..., y_{k-1}^w, y_k^w = \overline{b})$ be mutually-centered threshold values (w.l.o.g., $y_1^w \geq x_1^w$).
Let $\overline{x} = (\underline{a} = \overline{x_0}, \overline{x_1}, ..., \overline{x_{k-1}}, \overline{x_k} = \overline{b})$ and $\overline{y} = (\underline{a} = \overline{y_0}, \overline{y_1}, ..., \overline{y_{k-1}}, \overline{y_k} = \overline{b})$ be two threshold-value vectors for which the following constraints hold:

- $(\overline{x_1}, ..., \overline{x_{k-1}}, \overline{b})$ and $(\overline{y_1}, ..., \overline{y_{k-1}}, \overline{b})$ are mutually-centered vectors[13].

- $\overline{x_1} = v_0 \quad and \quad \overline{y_1} = \frac{1}{F_A(\overline{x_2})} \cdot \left( v_0 F_A(v_0) + \int_{\overline{x_1}}^{\overline{x_2}} v_A f_A(v_A) dv_A \right)$

The following theorem says that if the valuation of the seller for the item ($v_0$) is small enough (e.g., $\underline{a}$), the efficient mechanism is a priority game based on $x^w$ and $y^w$ (which are mutually centered). Otherwise, the optimal welfare can be achieved in a modified priority game based on $\overline{x}$ and $\overline{y}$.

**Theorem 3.5.** *For any pair of distribution functions of the bidders' valuations, and for any seller's valuation $v_0$ for the item, the mechanism $PG_k(x^w, y^w)$ **or** the mechanism $MPG_k(\overline{x}, \overline{y})$ achieves the optimal welfare (i.e., $w_{2,k}^{opt}$). In particular, $PG_k(x^w, y^w)$ achieves the optimal welfare when $v_0 = \underline{a}$.*

*Proof.* First, we prove that $PG_k(x^w, y^w)$ is optimal when $v_0 = \underline{a}$. According to Theorem 3.1 there is a pair of threshold values' vectors $x = (x_0, x_1, ..., x_k), y = (y_0, y_1, ..., y_k)$ such that $PG_k(x, y)$ achieves the optimal welfare. Note that $x_0 = y_0 = \underline{a}$ and $x_k = y_k = \overline{b}$, so we have $2(k-1)$ variables to optimize.

We will calculate the total expected welfare by summing first the expected welfare in the entries of the game matrix where $B$ wins the item, then summing the entries where $A$ is the winner.

$$w(g, s) = \sum_{i=1}^{k} (F_B(y_i) - F_B(y_{i-1})) \cdot (F_A(x_i) - F_A(x_0)) \cdot \frac{\int_{y_{i-1}}^{y_i} f_B(v_B) v_B dv_B}{F_B(y_i) - F_B(y_{i-1})}$$

---

13. Again, a unique solution exists when, w.l.o.g., $\overline{y_2} \geq \overline{x_2}$

$$+ \sum_{i=2}^{k} \left(F_A(x_i) - F_A(x_{i-1})\right) \cdot \left(F_B(y_{i-1}) - F_B(y_0)\right) \cdot \frac{\int_{x_{i-1}}^{x_i} f_A(v_A) v_A dv_A}{F_A(x_i) - F_A(x_{i-1})}$$

$$= \sum_{i=1}^{k} F_A(x_i) \cdot \int_{y_{i-1}}^{y_i} f_B(v_B) v_B dv_B + \sum_{i=2}^{k} F_B(y_{i-1}) \cdot \int_{x_{i-1}}^{x_i} f_A(v_A) v_A dv_A$$

We assume here that a probability density function exists for each bidder. Thus, we can express the partial derivatives with respect to all variables:

$$\left(w(g, s)\right)'_{x_i} = \left(\int_{y_{i-1}}^{y_i} f_B(v_B) v_B dv_B\right) \cdot f_A(x_i) + f_A(x_i) \cdot x_i \cdot F_B(y_{i-1}) - f_A(x_i) \cdot x_i \cdot F_B(y_i) = 0$$

$$\left(w(g, s)\right)'_{y_i} = \left(\int_{x_i}^{x_{i+1}} f_A(v_A) v_A dv_A\right) \cdot f_B(y_i) + f_B(y_i) \cdot y_i \cdot F_A(x_i) - f_B(y_i) \cdot y_i \cdot F_A(x_{i+1}) = 0$$

Rearranging the terms derives that $y_i = E_{v_A}(v_A \,|\, x_i \leq v_A \leq x_{i+1})$ and that $x_i = E_{v_B}(v_B \,|\, y_{i-1} \leq v_B \leq y_i)$ and therefore, $x, y$ should be mutually centered for optimal efficiency.

Now, we no longer assume $v_0 = \underline{a}$: According to Theorem 3.1, if the optimal welfare is not achieved in the priority game above, it will be achieved in a modified priority game. For some threshold values' vectors $x, y$, the expected welfare in $MPG_k(x, y)$ is given by the formula:

$$F_A(x_1) \cdot F_B(y_1) \cdot v_0 + F_A(x_1) \int_{y_1}^{\overline{b}} v_B f_B(v_B) dv_B + F_B(y_1) \int_{x_1}^{\overline{b}} v_A f_A(v_A) dv_A$$

$$+ \sum_{i=2}^{k} \left(F_A(x_i) - F_A(x_1)\right) \int_{y_{i-1}}^{y_i} v_B f_B(v_B) dv_B$$

$$+ \sum_{i=3}^{k} \left(F_B(y_{i-1}) - F_B(y_1)\right) \int_{x_{i-1}}^{x_i} v_A f_A(v_A) dv_A$$

First-order condition similarly derive the constraints on $x_1$ and $y_1$ given in the above definition of $\overline{x}, \overline{y}$, and that $(x_1, ..., x_{k-1}, x_k)$ and $(y_1, ..., y_{k-1}, y_k)$ should be mutually-centered[14]. $\qquad \square$

We demonstrate the characterization given above by showing an explicit solution for the case of uniformly-distributed valuations in $[0, 1]$.

**Corollary 3.6.** *When the bidders' valuations are distributed uniformly on $[0, 1]$ and $v_0 = 0$, the mechanism $PG_k(x, y)$ achieves the optimal welfare where*

$$x = (0, \frac{1}{2k-1}, \frac{3}{2k-1}, ..., \frac{2k-3}{2k-1}, 1) \ , \qquad y = (0, \frac{2}{2k-1}, \frac{4}{2k-1}, ..., \frac{2k-2}{2k-1}, 1)$$

---

14. The results are not surprising, since except for the case when one of the bidders bids 0, we have a priority game's allocation for which the optimal threshold values must be mutually centered (due to the first part of the proof).

*Proof.* According to Theorem 3.5 optimal welfare is achieved with $PG_k(x,y)$, when $x, y$ are mutually centered. With uniform distributions, this derives the following constraints, for which the given vectors $x, y$ are the unique solution: $\forall_{1 \le i \le k-1} \quad x_i = \frac{y_{i-1}+y_i}{2} \quad y_i = \frac{x_i+x_{i+1}}{2}$ To see how the above constraints are implied, note that the conditional expectation of the second player's value, given that his value is uniformly distributed between $y_{i-1}$ and $y_i$, is exactly $\frac{y_{i-1}+y_i}{2}$. $\qquad \blacksquare$

For example, when $k = 2$ we have the constraints $x_1 = \frac{0+y_1}{2}$ and $y_1 = \frac{x_1+1}{2}$, implying that $x_1 = 1/3$ and $y_1 = 2/3$ as in the optimal 1-bit mechanism from Example 1. The optimal mutually-centered thresholds for $k = 4$ are, for instance, $x = (0, \frac{1}{7}, \frac{3}{7}, \frac{5}{7}, 1)$ and $y = (0, \frac{2}{7}, \frac{4}{7}, \frac{6}{7}, 1)$.

### 3.3 Profit-Optimal 2-bidder Mechanisms with $k$ Possible Bids

Now, we present profit-maximizing 2-bidder mechanisms. Most results in the literature on profit-maximizing auctions assume that the distribution functions of the bidders' valuations are *regular* (as defined below). When the valuations of all bidders are distributed with the same regular distribution function, it is well known that Vickrey's 2nd-price auction, with an appropriately chosen reservation price, is profit-optimal (Vickrey, 1961; Myerson, 1981; Riley & Samuelson, 1981) with unbounded communication.

**Definition 14.** *(Myerson, 1981) Let $f$ be a probability density function, and let $F$ be its cumulative function. We say that $f$ is regular, if the function*

$$\widetilde{v}(v) = v - \frac{1-F(v)}{f(v)}$$

*is monotone, strictly increasing function of $v$. We call the function $\widetilde{v}(\cdot)$ the virtual valuation of the bidder.*

For example, when the bidders valuations are distributed uniformly on $[0, 1]$, a bidder with a valuation $v$ has a virtual valuation of $\widetilde{v}(v) = 2v - 1$.

**Definition 15.** *The virtual surplus in a game is the virtual valuation of the bidder (including the seller[15]) who receives the item.*

A key observation in the work of Myerson (1981), which we also use, is that in a Bayesian-Nash equilibrium, *the expected profit equals the expected virtual surplus* (in interim individually-rational equilibria where losing bidders are not getting any surplus). We use this property to reduce the profit-optimization problem to a welfare-optimization problem, for which we have already given a full solution. Myerson's observation was originally proved for direct-revelation mechanisms. We observe here that Myerson's observation also holds for auctions with bounded communication. That is, given a $k$-bid mechanism, the expected profit in every Bayesian-Nash equilibrium equals the expected virtual surplus.

**Proposition 3.7.** *Let $g \in G_{n,k}$ be a mechanism with a Bayesian Nash equilibrium $s = (s_1, ..., s_n)$ and interim individual rationality. Then, the revenue achieved by $s$ in $g$ is equal to the expected virtual surplus of $s$ in $g$.*

---

15. The seller's virtual valuation is defined to be his "original" valuation ($v_0$).

*Proof.* Consider the following direct-revelation mechanism $g_d$: each player $i$ bids her true valuation $v_i$. The mechanism calculates $s_i(v_i)$ for every $i$, and determines the allocation and payments according to $g$. An easy observation is that $g_d$ is incentive compatible (i.e., truthful bidding is a Bayesian-Nash equilibrium for the players) and interim individually rational. According to Myerson observation for direct revelation mechanisms, the expected revenue in $g_d$ is equal to the expected virtual surplus. However, for every combination of bids, both mechanism output identical allocations and payments. Thus, the expected revenue and the expected virtual surplus are equal in both mechanisms. $\qquad\square$

According to Theorem 3.5, the optimal *welfare* is achieved in either a priority game or a modified priority game. In a model where bidders consider their virtual valuations as their valuations, let $MPG(\widetilde{x}, \widetilde{y})$ or $PG(\overline{x}, \overline{y})$ be the mechanisms which are the candidates to achieve the optimal *welfare* (see Theorem 3.5 for a full characterization). Now, consider the same mechanisms, except each payment $\widetilde{c}$ in them is replaced by the respective "true" valuation $c = \widetilde{v}^{-1}(\widetilde{c})$ (i.e., $\widetilde{c} = \widetilde{v}(c)$ ). Denote these mechanisms by $PG_k(x^R, y^R)$, $MPG_k(x^r, y^r)$. These mechanisms achieve the optimal *profit* in our (original) model. Note that the distribution functions must be regular (but not necessarily identical) for this reduction to work.

**Theorem 3.8.** *When both bidders' valuations are distributed with regular distribution functions, the mechanism $MPG_k(x^r, y^r)$ **or** the mechanism $PG_k(x^R, y^R)$ (see definitions above) achieve the optimal expected profit among all profits achievable in an interim-IR Bayesian-Nash equilibrium of a mechanism in $G_{2,k}$ (i.e., $r_{2,k}^{opt}$).*

*Proof.* Consider the threshold-value vectors $(\widetilde{x}, \widetilde{y})$ and $(\overline{x}, \overline{y})$ defined above. The mechanism $MPG(\widetilde{x}, \widetilde{y})$ is efficient in the model where the bidders consider their virtual valuations as their valuation (the same proof holds if $PG(\overline{x}, \overline{y})$ is the efficient mechanism). The density function $f$ is regular, and therefore the virtual valuation $\widetilde{v}(\cdot)$ is strictly increasing. Thus, $MPG_k(x^r, y^r)$ (when the bidders use their original valuations) will have exactly the same allocation for every combination of bids as $MPG_k(\widetilde{x}, \widetilde{y})$ (when the bidders consider their virtual valuations as their valuations). We conclude that $MPG_k(x^r, y^r)$ achieves the optimal *expected virtual surplus* and thus also the optimal profit. $\qquad\square$

As in the case of welfare optimization, we give an explicit solution for the case of uniform distribution functions. This is a direct corollary of Theorem 3.8. Note that the optimal profit is achieved in a modified priority game. This holds since for the uniform distribution the bidders' expected virtual valuation is negative when they bid "0", so an efficient mechanism will not sell the item when all bidders bid "0".

**Corollary 3.9.** *When the bidders' valuations are distributed uniformly on $[0, 1]$ and $v_0 = 0$, the modified priority game $MPG_k(x, y)$ achieves the optimal expected profit among all the profits achievable in interim-IR Bayesian-Nash equilibria of mechanisms in $G_{2,k}$, where*

$$x = (0, \frac{1}{2}, \theta + \frac{1 \cdot (1 - \theta)}{2k - 3}, ..., \theta + \frac{(2k - 5) \cdot (1 - \theta)}{2k - 3}, 1)$$

$$y = (0, \theta, \theta + \frac{2 \cdot (1 - \theta)}{2k - 3}, ..., \theta + \frac{(2k - 4) \cdot (1 - \theta)}{2k - 3}, 1)$$

and $\theta = \frac{-2\alpha + \sqrt{1+3\alpha}}{2(1-\alpha)}$ for $\alpha = \frac{1}{(2k-3)^2}$ ($\theta = \frac{5}{8}$ when $k=2$).

## 4. Optimal Mechanisms for $n$ Bidders with Two Possible Bids

In this section we consider games among $n$ bidders where each bidder has 2 possible bids (i.e., they can send only 1 bit to the mechanism). We give the characterization of the optimal mechanisms for general distribution functions. The characterization of the optimal $n$-bidder mechanism with $k$ possible bids seems to be harder, and it remains an open question. The difficulty stems from the fact that the monotonicity of the allocation rule does not dictate the exact allocation rule in the general case. Rather, there are many possible allocation schemes that we cannot rule out before we know the strategies of the bidders.[16] Therefore, it seems that one should solve the involved combinatorial problem of finding the optimal allocation rule together with finding the optimal payments. Priority games with 2 possible bids per player can be interpreted as a sequence of take-it-or-leave-it offers; the player with the highest priority in this interpretation is the first player to be offered, if he accepts the offer (i.e., bids "1") he will receive it. See the work by Sandholm and Gilpin (2006) for the analysis of such take-it-or-leave-it mechanisms. Moreover, 1-bit priority games are actually a solution to a full-information version of the secretary problem (e.g., Gilbert & Mosteller, 1966; Ajtai, Megiddo, & Waarts, 2001). A decision maker meets the players (or potential secretaries to be hired), one after another, and should search for the player with the highest value. The decision maker knows the underlying probability distributions. The decision, however, should be made "online" – a player that does not receive the item upon arrival will never receive the item.

### 4.1 The Characterization of the Optimal Mechanisms

We first observe that priority games also maximize the welfare in $n$-bidder games with 2 possible bids. This is easier to see than in the $k$-possible-bids case. By Claim 3.2 in Theorem 3.1, the bidders will use threshold strategies. An efficient mechanism will allocate the item, for each combination of bids $\overrightarrow{b}$, to the bidder with the highest expected welfare when he bids $b_i$. Given that the distributions are i.i.d, if $x_i \geq x_j$ then $E\left(v_i \mid v_i \in [x_i, \overline{b}]\right) \geq E\left(v_j \mid v_j \in [x_j, \overline{b}]\right)$ and $E\left(v_i \mid v_i \in [\underline{a}, x_i]\right) \geq E\left(v_j \mid v_j \in [\underline{a}, x_j]\right)$. Therefore, ties will be broken according to the order of the thresholds. If the seller's reservation price $v_0$ is high enough, the efficient mechanism will be a modified priority game.[17]

We now show the characterization of the optimal thresholds for the priority games. We show that the optimal mechanisms use fully discriminatory payments: the bidder with the highest priority in the priority game pays the highest payment when she wins, and so forth. The optimal modified priority game is given by a simple recursive formula. When the seller allocates the item for when all bids are zero, the constraints become cyclic.

---

16. Consider, for example, a 3-bidder 3-bid priority game, where the item is allocated to the player with the second-highest priority when all the players bid their highest bid. This mechanism is also monotone with no identical actions for the players.

17. To see this, we must note that in an efficient mechanism the seller will never keep the item when one of the bidder bids 1 (then, a threshold higher than $v_0$ for this bidder will gain a higher welfare).

Let $\overrightarrow{x} = (x_1, ..., x_n)$ and $\overrightarrow{y} = (y_1, ..., y_n)$ be the profiles of threshold values for the $n$ bidders such that the following constraints hold:

$$x_1 = E\left(v \,\middle|\, \underline{a} \leq v \leq x_n\right) \tag{1}$$

$$\forall_{1 \leq m \leq n-2} \quad x_{m+1} = (1 - F(x_m)) \cdot E\left(v \,\middle|\, v \in [x_m, \overline{b}]\right) + F(x_m) \cdot x_m \tag{2}$$

$$x_n = \frac{\sum_{i=1}^{n-1}(\prod_{j=i+1}^{n-1} F(x_j))(1 - F(x_i)) \, E\left(v \,\middle|\, v \in [x_i, \overline{b}]\right)}{1 - \prod_{i=1}^{n-1} F(x_i)} \tag{3}$$

$$y_1 = v_0 \tag{4}$$

$$\forall_{1 \leq m \leq n-2} \quad y_{m+1} = (1 - F(y_m)) \cdot E\left(v \,\middle|\, v \in [y_m, \overline{b}]\right) + F(y_m) \cdot y_m \tag{5}$$

We will now prove that either the mechanism $PG_2(\overrightarrow{x})$ or the mechanism $MPG_2(\overrightarrow{y})$ achieve the optimal welfare. As the thresholds description shows, the thresholds for the modified priority game (i.e., when the seller keeps the item when all bids are zero) are defined by a simple, easy-to-compute recursive formula. The optimality of these thresholds can be shown by the following intuitive argument: Consider a new bidder $i$ that joins a set of $i-1$ bidders. An efficient auction will allocate the item to bidder $i$ if and only if his value is greater than the optimal welfare achievable from the first $i-1$ bidders. Therefore, the threshold for each bidder will equal the optimal welfare gained from the preceding bidders; and indeed, with probability of $1 - F(y_{i-1})$, bidder $i$'s valuation will be greater than the expected welfare attained from the other bidders ($y_{i-1}$) and his average contribution will be $E\left(v \,\middle|\, v \in [y_m, \overline{b}]\right)$); with probability of $F(y_{i-1})$ he will not contribute to the optimal welfare which remains $y_{i-1}$. This intuition shows why the profit-maximizing thresholds above (Equations 4,5) are independent of the number of players, what enables an "online" implementation of the profit-maximizing auctions.

**Theorem 4.1.** *When the bidders' valuations are distributed with the same distribution function, the mechanism $PG_2(\overrightarrow{x})$ or the mechanism $MPG_2(\overrightarrow{y})$ achieves the optimal expected welfare. In particular, when $v_0 = \underline{a}$, $PG_2(\overrightarrow{x})$ is the efficient mechanism.*

*Proof.* We already observed that there exists a priority game that achieves the optimal welfare with threshold strategies. Consider a priority game among $n$ bidders, indexed by their priorities (i.e., $1 \prec 2... \prec n$). Every bidder wins the item if he bids 1 and all the bidders *with higher priorities* bid 0. Thus, the probability that bidder $i$ wins is $\left(\prod_{j=i+1}^{n} F(x_j)\right) \cdot (1 - F(x_i))$. When all bidders bid 0, either bidder $n$ wins or the seller keeps the item for herself. The expected welfare from this game, where the bidders use threshold strategies $x_1, ..., x_n$ is:

$$w(g,s) = \sum_{i=1}^{n} \left(\prod_{j=i+1}^{n} F(x_j)\right)(1 - F(x_i)) \frac{\int_{x_i}^{\overline{b}} f(v_i) v_i dv_i}{(1 - F(x_i))} + \left(\prod_{i=1}^{n} F(x_i)\right) E_0$$

Where $E_0 = E(v_n | v_n \in [\underline{a}, x_n])$ in the priority game and $E_o = v_0$ in a modified priority game (the second term relates to the case when all the bidders bid 0). For maximum, the partial derivatives with respect to $x_1, ..., x_n$ should equal zero, resulting a characterization of the optimal solution.

251

For bidders $1 \leq m \leq n-1$ we get that $x_m$ equals (both in the priority game and in the modified priority game):

$$\sum_{i=1}^{m-1} \left( \prod_{j=i+1}^{m-1} F(x_j) \right) (1 - F(x_i)) E\left( v_i \,|\, v_i \in [x_i, \overline{b}] \right) + \left( \prod_{i=1, i \neq m}^{m-1} F(x_i) \right) E\left( v_n \,|\, v_n \in [\underline{a}, x_n] \right)$$

The recursive formula is reached by calculating $x_{m+1} - x_m$, from which Equations 2 and 5 follow. For bidder $n$ in the priority game the first order conditions yield the constraint in Equation 3. When $m = 1$, we have $x_1 = E\left( v_n \,|\, \underline{a} \leq v_n \leq x_n \right)$ (in the priority game) and $x_1 = v_0$ (in the modified priority game). $\qquad \square$

As in Section 3, we characterize the profit optimal mechanism by a reduction to the welfare optimizing problem. Again, the reduction can be performed only for regular distributions. Consider the model where bidders take their virtual valuations as their valuations. Let $PG_2(\widetilde{u})$ or $MPG_2(\widetilde{z})$ be the mechanisms that achieve the optimal *welfare* in this model (see Theorem 4.1 above). Let $PG_2(u)$ and $MPG_2(z)$ be similar mechanisms respectively, except each payment $\widetilde{c}$ is replaced with its respective "original" valuation $c = \widetilde{v}^{-1}(\widetilde{c})$.

**Theorem 4.2.** *When the bidders' valuations are distributed with the same regular distribution function, the mechanism $PG_2(u)$ or the mechanism $MPG_2(z)$ achieves the optimal expected profit among all the profits achievable with a Bayesian-Nash equilibrium and interim IR.*

*Proof.* This is a corollary of Theorem 4.1. The reduction is done as in Theorem 3.8, and it is possible due to the regularity of the distribution function. $\qquad \square$

Again, the optimal thresholds for the modified priority game can be given by a simple recursive formula with an intuitive meaning. The recursion is identical to the welfare optimizing formula (Equation 2), and the only difference is in the value of the first threshold that should hold $y_1 = \widetilde{v}^{-1}(v_0)$. The intuition is that given that the best revenue achievable from the first $i-1$ bidders is $y_{i-1}$, with probability $F(y_{i-1})$ a new player $i$ will not be able to pay a higher price (due to the individual-rationality restriction) and therefore the optimal revenue remains $y_{i-1}$. When his value is greater than $y_{i-1}$, he cannot be charged more than his average value ($E(v|y_{i-1} \leq v \leq \overline{b})$).

Now, we give explicit solutions for the uniform distribution on the support $[0, 1]$. The following recursive constraints characterize the efficient and profit-optimal mechanisms – these are the constraints given in Theorems 4.1 and 4.2 for uniform distributions.

Let $(x_1, ..., x_n) \in [0, 1]^n$ be threshold values for which the following constraints hold:

$$x_1 \;=\; \frac{x_n}{2} \tag{6}$$

$$\forall m \in \{1, ..., n-2\} \quad x_{m+1} \;=\; \frac{1}{2} + \frac{x_m^2}{2} \tag{7}$$

$$x_n \;=\; \frac{\sum_{i=1}^{n-1} \left( \prod_{j=i+1}^{n-1} x_j \right) \left( 1 - x_i^2 \right)}{2 \left( 1 - \prod_{i=1}^{n-1} x_i \right)} \tag{8}$$

Let $y = (y_1, ..., y_n) \in [0, 1]^n$ be threshold values where $y_1 = \frac{1}{2}$ and:

$$\forall m \in \{1, ..., n-2\} \quad y_{m+1} = \frac{1}{2} + \frac{y_m^2}{2}$$

**Corollary 4.3.** *Consider the threshold values $\overrightarrow{x} = (x_1, ..., x_n)$ and $\overrightarrow{y} = (y_1, ..., y_n)$ defined above. When the bidders' valuations are distributed uniformly in $[0, 1]$ and $v_0 = 0$, $PG_2(\overrightarrow{x})$ achieves the optimal welfare and $MPG_2(\overrightarrow{y})$ achieves optimal profit.*

For example, when $n = 5$ we have $y = (0.5, 0.625, 0.695, 0.741, 0.775)$. The above description is simpler than any closed-form formulae we could find.

## 5. Asymptotic Analysis of the Welfare and Profit Losses

In this section, we measure the performance of the optimal mechanisms presented in earlier sections. Although we did not present a characterization of the optimal mechanisms in the general model of $k$ possible bids and $n$ bidders, we present here mechanisms for this general case that are *asymptotically* optimal. For simplicity, we assume that the valuations' range is $[0, 1]$ (all the results apply for a general range $[\underline{a}, \overline{b}]$ which only changes the constants in our analysis).

We analyze the welfare loss (Subsection 5.1), the profit loss (Subsection 5.2), and finally, in Subsection 5.3 we measure the profit loss and the welfare loss in 1-bit mechanism with $n$ bidders. All the result are asymptotic with respect to the amount of the communication, except in Section 5.3 where it is with respect to the number of bidders.

### 5.1 Asymptotic Bounds on the Welfare Loss

The next theorem shows that no matter how the bidders' valuations are distributed, we can always construct mechanisms such that the welfare loss they incur diminishes quadratically in $k$. This is true for any number of bidders we fix (when $k > 2n$). In particular, the efficient mechanism presented in Theorem 3.5 incurs a welfare loss of $O(\frac{1}{k^2})$. The intuition behind the proof: given the distribution functions of the bidders, we construct a certain threshold strategy, which will be dominant for all bidders. When using this strategy, each bidder will bid any bid $i$ with probability smaller than $\frac{1}{k}$. This way, the probability that a welfare loss may occur is $O(\frac{1}{k})$ (for two players, for instance, a welfare loss will be possible only on the diagonal of the game's matrix). The average welfare loss will also be $O(\frac{1}{k})$, resulting in a total expected loss of $O(\frac{1}{k^2})$.

**Theorem 5.1.** *For any (fixed) number of bidders $n$, and for any set of distribution functions of the bidders' valuations, there exist a set of mechanisms $g_k \in G_{n,k}$ ($k = 2n+1, 2n+2, ...$), that incur an expected welfare loss of $O(\frac{1}{k^2})$). These results are implemented in dominant strategies with ex-post individual rationality.*

The requirement that $k > 2n$ (here and in Proposition 5.3 below) is due to the construction of the symmetric mechanism in the following proof. These results hold even without this requirement, as shown by an asymmetric construction for a more general setting in the work of Blumrosen and Feldman (2006).

*Proof.* The proof's idea: we construct a priority game in which all bidders have the same dominant threshold strategy, such that the probability for a bidder to bid each bid is smaller than $\frac{n}{k}$. This is done by dividing the density functions of all the bidders to $\frac{k}{n}$ intervals with equal mass, then combining these thresholds to a vector of $k$ threshold values. Because the bidders use the same threshold strategy, a welfare loss is possible only when more than one bidder bids the highest bid. This observation leads to the upper bound.

Let $\alpha_1, ..., \alpha_n$ be integers such that $\sum_{i=1}^{n} \alpha_i = k - 2$, and for every $i$, $\alpha_i \geq \lfloor \frac{k}{n} \rfloor - 1$ (clearly such numbers exist). For every bidder $i$, let $Y^i = (y_1^i, ..., y_{\alpha_i}^i)$ be a set of threshold values that divide her distribution function $f_i$ to $\alpha_i + 1$ segments with the same mass (when $y_0^i = 0, y_{\alpha_i + 1}^i = 1$), i.e., for every bid $j$, $F_i(y_{j+1}) - F_i(y_j) = \frac{1}{\alpha_i + 1}$.

Let $X = \{\bigcup_{i=1}^{n} Y^i\} \bigcup \{v_0\}$, $|X| = k - 1$, be the union of all the threshold values (we add arbitrary threshold values if the size of $X$ is smaller than $k - 1$). Let $x = (0, x_1, ..., x_{k-1}, 1)$ be a threshold-value vector created by ordering the threshold values in $X$ from smallest to largest. Now, consider the $n$-bidder mechanism $MPG_k(\tilde{t})$ where $\tilde{t} = (x, .., x)$. The threshold strategy based on $x$ is dominant for all the bidders, with ex-post IR. By the construction of the sets $Y^1, ..., Y^n$, every bidder will bid any particular bid w.p. $\leq \frac{2n}{k}$.[18]

Next, we will bound the welfare loss. We divide the possible cases according to the number of bidders who bid the highest bid. Since all the bidders use the same threshold strategy, if only one bidder bids the highest bid, no welfare loss is incurred (he will definitely have the highest valuation). If more than 1 bidder bid the highest bid $i$, the expected welfare loss will not exceed $x_{i+1} - x_i$. For a set of bidders $T \subseteq N$, denote the probability that all the bidders in $T$ bid $i$ by $Pr(T = i)$, and the probability that all the bidders not in $T$ have bids smaller than $i$ by $Pr(N \setminus T < i)$. Thus, the expected welfare loss is smaller than (when $2n < k$):

$$\sum_{j=2}^{n} \sum_{T \subseteq N, \, |T|=j} \sum_{i=1}^{k} Pr(T = i) Pr(N \setminus T < i) \, (x_{i+1} - x_i)$$

$$\leq \sum_{j=2}^{n} \sum_{T \subseteq N, \, |T|=j} \sum_{i=1}^{k} Pr(T = i) \, (x_{i+1} - x_i)$$

$$\leq \sum_{j=2}^{n} \sum_{1}^{\binom{n}{j}} \sum_{i=1}^{k} \left( \frac{2n}{k} \right)^j (x_{i+1} - x_i) \quad = \quad \sum_{j=2}^{n} \binom{n}{j} \left( \frac{2n}{k} \right)^j \quad < \quad 2^n \cdot 4n^2 \cdot \frac{1}{k^2}$$

When the valuations of all the bidders are smaller than $v_0$, there is no welfare loss (it is easy to see that we can assume, w.l.o.g., that $x_1 = v_0$). Note that despite the coefficient of $\frac{1}{k^2}$ is exponential in $n$, we consider it as a constant because $n$ is fixed. For Example, when $n = 2$ a similar proof shows a welfare loss smaller than $\frac{8}{k^2}$ (when $k > 3$). $\qquad \square$

Asymptotic quadratic bounds were also given by Wilson (1989), which studied similar settings regarding the effect of discrete priority classes of customers. In the work of Wilson the uncertainty was about the supply, while in this paper the demand is uncertain as well. Both results are illustrations for the idea that the deadweight loss is second order in the price distortion. (The price distortion in our model is the maximum difference between the prices that different bidders are facing for the item given the others' bids, and it can be bounded above by $\frac{1}{k}$.) Indeed, a small price distortion ensures both that the probability of an inefficient allocation is small and that the inefficiency is small when it does occur.

---

18. For every bidder $i$, and every bid $j$, $F_i(x_{j+1}) - F_i(x_j) \leq \frac{1}{\lfloor \frac{k}{n} \rfloor} \leq \frac{2n}{k}$

Theorem 5.1 is related to proposition 4 in the paper by Nisan and Segal (2006). Nisan and Segal showed that discretizing an exactly efficient continuous protocol communicating $d$ real numbers yields a "truly polynomial" approximation scheme that is proportional to $d$ (i.e., for any $\epsilon > 0$ we can realize an approximation factor of $1 - \epsilon$ using a number of bits which is polynomial in $log(\epsilon^{-1})$ ). Here, we discretize a continuous efficient auction (e.g., first-price auction), where $d$ is the number of bidders. Discretization then achieves an approximation error that is exponential in the (minus) number of bits sent per bidder, i.e., asymptotically proportional to $\frac{1}{k}$. However, here we care about average-case approximation which is even closer, because worst-case approximation within an error of $\epsilon$ ensures an average case approximation within $\epsilon^2$ (the probability that an error is made is itself in the order of $\epsilon$).

We now show that the asymptotic upper bound above is tight, i.e., for some distribution functions (and in particular, for the uniform distribution) the minimal welfare loss is exactly proportional to $\frac{1}{k^2}$. We show this for any constant number of bidders.

**Theorem 5.2.** *Assume that the bidders' valuations are uniformly distributed and that $v_0 = 0$. Then, the efficient 2-bidder mechanism $PG_k(x, y)$ described in Corollary 3.6 incurs a welfare loss of exactly $\frac{1}{6 \cdot (2k-1)^2}$. Moreover, for any (fixed) number of bidders $n$ and for any $v_0$, there exists a positive constant $c$ such that **any** mechanism $g \in G_{n,k}$ incurs a welfare loss $\geq c \cdot \frac{1}{k^2}$.*

*Proof.* We first prove the first part of the theorem, regarding 2-bidder mechanisms. Note that the given mechanism can make non-optimal allocation only for combinations of bids that are on the diagonal or on the lower secondary diagonal in the matrix representation of the 2-bidder game (i.e., when $b_A = b_B$ or when $b_A = b_B + 1$). For such bids $(i, j)$, the overlapping segment of $[x_i, x_{i+1}]$ and $[y_j, y_{j+1}]$ is of size $\frac{1}{2k-1}$. Given such vector of bids $(i, j)$, if one of the valuations is not in this overlapping segment, the allocation is optimal (note that we allocate the item to $B$ on the main diagonal, and to $A$ on the secondary diagonal). The probability that both valuation are in this overlapping range is $\frac{1}{(2k-1)^2}$. The expected valuation in our priority game (when both valuation are in this overlapping segment) is exactly in the middle of this segment. The expected valuation in the optimal auction (with unbounded communications), restricted to this overlapping interval, will be in the $\frac{2}{3}$ point of this range. Thus, the welfare loss is $\frac{1}{6}$ of the segment, i.e., $\frac{1}{6} \cdot \frac{1}{2k-1}$. Thus, for every vector of bids on the main diagonal or on the secondary-diagonal the expected welfare loss is $\frac{1}{6} \frac{1}{(2k-1)^3}$. There are $(2k-1)$ such vector of bids, thus the total welfare loss is exactly $\frac{1}{6(2k-1)^2}$.

A similar argument shows that even when the seller's valuation $v_0$ is non zero, the welfare loss is asymptotically greater than $\frac{1}{(2k-1)^2}$: let $z_1, ..., z_m$ be the sizes of the overlapping segments (only when the valuations of both bidders are greater than $v_0$). Clearly, $m \leq 2k - 1$ and $\sum_{i=1}^{m} z_i \leq 1$. Then, the welfare loss from the game is at least [19]:

$$(1-v_0)^2 \cdot \sum_{i=1}^{m} z_i^2 \cdot \frac{z_i}{6} = \frac{(1-v_0)^2}{6} \cdot \sum_{i=1}^{m} z_i^3 \geq \frac{(1-v_0)^2}{6} \frac{2k-1}{(2k-1)^3} \geq \frac{(1-v_0)^2}{6} \frac{1}{(2k-1)^2}$$

The proof of the second statement is easily derived: Consider only the case where bidders 1 and 2 have valuations above $\frac{1}{2}$, and the rest of the bidders have valuations below $\frac{1}{2}$. This occurs with the constant probability of $\frac{1}{2^n}$. The best a mechanism can do is to always allocate the item to one of 1 or 2. But due to the first part of the theorem, in any 2-bidder mechanism a welfare loss of proportional to $\frac{1}{k^2}$ will be incurred (the fact that the valuation range is $[\frac{1}{2}, 1]$ and not $[0, 1]$ only changes the constant $c$). This will hold for any opportunity cost $v_0$ of the seller. Thus, any mechanism will incur a welfare loss of $\Omega(\frac{1}{k^2})$. □

Note that the same asymptotic results hold even if we restrict attention to symmetric mechanisms. Actually, we prove the upper bound in Theorem 5.1 by constructing a symmetric mechanism (we can allocate the item to all the bidders who bid the highest bid with equal probabilities). However, asymmetric mechanisms do incur a strictly smaller welfare loss than symmetric mechanisms. For example, when the valuations are distributed uniformly, the optimal welfare loss is $\frac{1}{6(2k-1)^2}$ (by Theorem 5.2) compared with an optimal welfare loss of $\frac{1}{6k^2}$ attained by symmetric mechanisms[20] (i.e., the welfare loss in asymmetric mechanisms is about 4 times better). This observation is interesting in light of the results of Harstad and Rothkopf (1994) and Wilson (1989). Harstad and Rothkopf studied symmetric English auctions, and analyzed the optimal price-jumps in such auctions. Our results show that non-anonymous prices (i.e., different jumps for each bidder) can achieve better results than symmetric (or anonymous) jumps. We also characterize the optimal price-jumps for such auctions (mutually centered threshold values). Wilson also studied only symmetric priority classes in his model, and also gives a convergence rate of $\frac{1}{n^2}$ for the efficiency loss (where $n$ is the number of priority classes). We show that asymmetric mechanisms can incur smaller efficiency loss, although the asymptotic convergence rate is the same.

One obvious drawback of our characterization of the optimal mechanisms is that the design is not "detail-free" (as in Wilson's doctrine) – we must know the priors of the bidders for designing the mechanisms. But can we design a mechanism that regardless of the distribution functions, will always incur a low welfare loss? The answer is that we can, but they will not be as efficient as in the commonly-known priors case. We observe that a simple, symmetric mechanism that use equally spaced thresholds (i.e., $PG_k(x, ..., x)$, $x = (0, \frac{1}{k}, \frac{2}{k}, ..., \frac{k-1}{k}, 1)$ ), incurs a welfare loss not greater than $\frac{1}{k}$ for all possible distribution functions. It is not hard to verify that this is actually the best that can be done by "detail-free" mechanism: for any mechanism there exist distribution functions for which the expected welfare loss is at least in order of $\frac{1}{k}$. For severely low communication, the difference between the "detail-free" mechanisms and prior-aware mechanisms (with loss of $O(\frac{1}{k^2})$ ) may be substantial. Note that without communication constraints, socially-efficient results can be achieved by "detail-free" mechanisms – second-price (Vickrey) auctions.

---

19. In the left inequality we use the fact that when $z = (z_1, ..., z_m)$ is in the $m$'th dimensional simplex, $\sum_{i=1}^{m} z_i^3 \geq \frac{1}{m^2}$.

20. It is easy to show that efficient symmetric mechanisms are similar to priority games, except the item is allocated with equal probabilities in cases of ties. The thresholds of the bidders simply divide the valuations' range to identical segments. Then, it is straightforward to show that the welfare loss is exactly $\frac{1}{6k^2}$.

## 5.2 Asymptotic Bounds on the Profit Loss

As done in Theorem 3.8, the profit optimization problem can be reduced to a welfare-optimization problem by maximizing the expected virtual surplus.

**Proposition 5.3.** *Assume that the bidders' valuations are distributed with regular distribution functions. Then, for any number of bidders $n$, there exist a set of mechanisms $g_k \in G_{n,k}$ ($k = 2n + 1, 2n + 2, ...$) that incur a profit loss of $O(\frac{1}{k^2})$. The profit loss is compared with the optimal, individually-rational mechanism that is unconstrained in communication.*

*Proof.* Consider the model where bidders consider their virtual valuations $\widetilde{v_i}(v_i)$ as their valuations. As the range of the valuations in this model, we take the union of the ranges of all the bidders' virtual valuations. Denote this range as $[\alpha, \beta]$. Let $\widetilde{g} \in G_{n,k}$ be the mechanism that achieves the maximal *welfare* in this model. Due to Theorem 5.1, $\widetilde{g}$ incurs a welfare loss smaller than $c \cdot \frac{1}{k^2}$, for some positive constant $c$ (the constant takes into account the size of the virtual valuations' range $\beta - \alpha$). Let $g$ be the mechanism with the same allocation as in $\widetilde{g}$, only each payment $\widetilde{q_i}$ for bidder $i$ in $\widetilde{g}$ is replaced with $q_i = \widetilde{v_i}^{-1}(\widetilde{q_i})$ in $g$, i.e., $\widetilde{q_i} = \widetilde{v_i}(q_i)$. Since each $\widetilde{v_i}$ is non-decreasing (by their regularity), the allocation rules in $g$ and $\widetilde{g}$ are identical for every bids' combination. Thus, $g$ achieves the maximal expected virtual surplus, and the loss of expected virtual surplus is smaller than $c \cdot \frac{1}{k^2}$. The proposition follows. □

Again, this upper bound is asymptotically tight: with the uniform distribution, any mechanism incurs a profit loss of $\Omega(\frac{1}{k^2})$. This result is derived from Theorem 5.2 using similar arguments as in Proposition 5.3.

**Proposition 5.4.** *Assume that the bidders' valuations are distributed uniformly. Then, for any (fixed) number of bidders $n$, there exists a positive constant $c$ such that **any** mechanism $g \in G_{n,k}$ incurs a profit loss $\geq c \cdot \frac{1}{k^2}$.*

So far, we assumed that the bidders' valuations are drawn from statistically independent distributions. We now point out that the relaxation to general joint distributions is non-interesting in our model. Specifically, we can show that the trivial priority game for which all the bidders use the threshold strategy based on the vector $x = (0, \frac{1}{k}, \frac{2}{k}, ..., \frac{k-1}{k}, 1)$ always incurs an expected welfare loss smaller than $\frac{1}{k}$, and no mechanism can do asymptotically better. In other words, there exists some joint distribution function for which *any* mechanism incurs a welfare loss proportional to $\frac{1}{k}$.

## 5.3 Asymptotic Bounds for a Growing Number of Bidders

In this subsection, we fix the size of communication allowed (to two possible bids), and we show asymptotic bounds as a function of the number of bidders rather than the amount of communication. Unfortunately, we have been able to prove such bounds only for the uniform distribution.

When we restrict our attention to *symmetric* mechanisms, the solution is simple. Using the threshold $x = n^{-\frac{1}{n-1}}$ (for all bidders) achieves the maximal expected welfare, and we have the exact formula showing that the optimal welfare loss is $O(\frac{\log n}{n})$.[21]

We now show that optimal *asymmetric* mechanisms incur asymptotically smaller welfare and profit losses of $O(\frac{1}{n})$. These mechanisms fully discriminate between the agents.

**Theorem 5.5.** *Consider the mechanisms $PG_2(\overrightarrow{x})$ and $MPG_2(\overrightarrow{y})$ described in Corollary 4.3 (in Section 4.1). When the bidders' valuations are distributed uniformly, both the welfare loss in $PG_2(\overrightarrow{x})$ and the profit loss in $MPG_2(\overrightarrow{y})$ are smaller than $\leq \frac{9}{n}$.*

*Proof.* Let $x$ be the *revenue*-optimizing thresholds from Corollary 4.3. We will bound the *welfare* loss in $PG_2(x)$, the efficient mechanism will incur even a smaller loss. We assume, w.l.o.g. that in $g$, bidders are indexed according to their priorities (i.e., $1 \prec 2... \prec n$ ). When a bidder wins after bidding "1", the maximal welfare loss is $1 - x_i$. When all bidders bid "0", we use the trivial upper bound of 1 for the welfare loss . Therefore, we can bound the welfare loss with:

$$\sum_{i=1}^{n} \left( \prod_{j=i+1}^{n} x_j \right) (1-x_i)(1-x_i) + \prod_{i=1}^{n} x_i \qquad (9)$$

The following two claims can be easily verified by induction:

**Claim 5.6.** $\forall_n \quad 1 - x_n \leq \frac{2}{n}$

**Claim 5.7.** $\forall_{n \geq 15} \quad x_n \leq \frac{2n-3}{2n}$

Now, we prove by induction on $n$ that the first summand in Equation 9 is $\leq \frac{8}{n}$. Denote this first term by $\overline{wl_n}$. Note that $\overline{wl_{n+1}} = (1-x_{n+1})^2 + x_{n+1}\overline{wl_n}$. Assuming that $\overline{wl_n} \leq \frac{8}{n}$, and using the two claims above, it is easy to prove that $\overline{wl_{n+1}} \leq \frac{8}{n+1}$ for $n > 14$. (the reader can verify that this also holds for $n \leq 14$.)

Next, we prove (again by induction on $n$) that the second expression is smaller than $\frac{1}{n}$. We assume $\prod_{i=1}^{n} x_i \leq \frac{1}{n}$ and prove that $\prod_{i=1}^{n+1} x_i \leq \frac{1}{n+1}$ (using Claim 5.7 ) :

$$\prod_{i=1}^{n+1} x_i = x_{n+1} \prod_{i=1}^{n} x_i \leq x_{n+1}\frac{1}{n} \leq \frac{2n-1}{2n+2} \cdot \frac{1}{n} < \frac{2n-1}{2n+2} \cdot \frac{1}{n} + \frac{1}{2n(n+1)} = \frac{1}{n+1}$$

Thus, the expected welfare loss is smaller than $\frac{8}{n} + \frac{1}{n} = \frac{9}{n}$

The statement about the profit loss can be derived from the result about the welfare loss (again, by reducing profit optimization to welfare optimization). Nevertheless, a direct proof is easy given the above claims: with the same thresholds $x$ from above, the profit loss is bounded from above by $\sum_{i=1}^{n} \left( \prod_{j=i+1}^{n} x_j \right) \cdot (1 - x_i) \cdot (1 - x_i)$ that was proved to be smaller than $\frac{8}{n}$.[22] $\qquad\square$

---

21. The expected welfare then is given by: $x^n \cdot \frac{x}{2} + (1 - x^n) \cdot \frac{1+x}{2}$. A maximum is achieved (first order conditions) with: $x = n^{-\frac{1}{n-1}}$. The welfare loss is thus: $\frac{n}{n+1} - \frac{1}{2}\left(1 - n^{-\frac{1}{n-1}}(\frac{1}{n} - 1)\right)$ ($\frac{n}{n+1}$ is the maximal welfare with unbounded communication). It is easy to see that if $1 - \left(\frac{1}{n}\right)^{\frac{1}{n}}$ converges to $\frac{\log n}{n}$ then the welfare loss also converges to $\frac{\log n}{n}$. And indeed, $1 - \left(\frac{1}{n}\right)^{\frac{1}{n}} = 1 - e^{-\frac{\log n}{n}} \approx \frac{\log n}{n}$ (since $1 - e^{-x} \approx x$ for small $x$'s).

| B ╲ A | 0 | 1 |
|---|---|---|
| 0 | $A, 0$ | $B, \frac{1}{4}$ |
| 1 | $A, \frac{1}{3}$ | $B, \frac{3}{4}$ |

Figure 3: ($h_1$) This sequential game (when $A$ bids first) attains a higher expected welfare than any simultaneous mechanism with the same communication requirement (2 bits). This outcome is achieved with Bayesian-Nash equilibrium.

## 6. Sequential Auctions

In sequential mechanisms, bidders split their bids into smaller messages and send them in an alternating order. In this section, we show that sequential mechanisms can achieve better results. However, the additional gain (in the amount of communication) is only up to a linear factor in the number of bidders.

A *sequential* mechanism is a mechanism in which each bidder may send several separate messages, in some order (not necessarily in a round-robin fashion). At each stage, each bidder knows what messages the other bidders have sent so far. After all the messages are sent, the mechanism determines the allocation and payments. We study a general framework where the auctioneer can adaptively determine the order of the messages and their sizes according to the message history. The auctioneer can also use randomization in its decisions. We measure the communication volume in a mechanism by the number of bits actually transmitted.

**Definition 16.** *The communication requirement of the mechanism is the maximal amount of bits which may be transmitted by the bidders in this mechanism.*

A strategy for a bidder in a sequential mechanism is a *threshold strategy* if in each stage $i$ of the game the bidder determines the message she sends by comparing her valuation to some threshold values $x_1, ... x_{\alpha_i}$ (where this bidder has $\alpha_i + 1$ possible bids in stage $i$).

**Example 2.** *The following sequential mechanism has a communication requirement of 2 (see Figure 3 ): Alice sends one bit to the mechanism first. Bob, knowing Alice's bid, also sends one bit. When Alice bids 0: Bob wins if he bids 1 and pays $\frac{1}{4}$; If he bids zero Alice wins and pays zero. When Alice bids 1: Bob also wins when he bids 1, but now he pays $\frac{3}{4}$; If he bids zero, Alice wins again, but now she pays $\frac{1}{3}$.*

*It is easy to see that this mechanism has a Bayesian-Nash equilibrium[23] that achieves an expected welfare of 0.653. We saw that the efficient* simultaneous *mechanism with a*

---

22. In the priority games based on the thresholds $\overrightarrow{y}$, if bidder $i$ wins the item, he pays $y_i$. Thus, the maximal profit loss when bidder $i$ wins is $1 - y_i$.

23. The following strategies are in Bayesian-Nash equilibrium: Alice uses the threshold $\frac{1}{2}$, and Bob uses the threshold $\frac{1}{4}$ when Alice bids 0 and $\frac{3}{4}$ when Alice bids 1.

*communication requirement of 2 bits is 0.648 (see Section 1). We conclude that sequential mechanisms can gain more efficiency than simultaneous mechanisms.*

Note that throughout the paper we searched for optimal mechanisms among all the mechanisms with Bayesian-Nash equilibria, but we managed to implement this optimum in dominant strategy. In sequential mechanisms it is less likely to find dominant-strategy implementations, thus our above example uses Bayesian-Nash implementation. Our result below, however, do not assume any particular equilibrium concept in the sequential mechanisms.

How significant is the extra gain from sequential mechanisms over simultaneous mechanisms? The following theorem states for every sequential mechanism with a communication requirement of $m$ there exists a simultaneous mechanism that achieves at least the same welfare with a communication requirement of $nm$ (where $n$ is the number of bidders)[24]. Note that in general (e.g., Kushilevitz & Nisan, 1997), multi-round protocols can reduce the communication by an exponential factor. We observe that the gain from sequential mechanism is actually even smaller. In many environments, all messages are sent to a centralized authority (auctioneer); therefore, extra bits of communication will be required to inform the bidders about the previous messages of the other bidders. The following theorem holds for any order of transmission and any size of the sub-messages, even if these values are adaptively determined according to previous messages.

The goal of this section is to show that the gain from sequential auctions, compared to simultaneous auctions, is mild. We do not offer a comprehensive analysis of this case, not present welfare-maximizing and revenue-maximizing auctions. Several recent papers studied different aspects of sequential auctions with similar constraints. Sandholm and Gilpin (2006) analyzed sequential auctions designed as sequences of take-it-or-leave-it offers. A paper by Kress and Boutilier (2004) studied sequential single-item auctions with discrete price increment, where information can be used in subsequent stages. The work of Parkes (2006) studied information elicitation in simultaneous and sequential auctions when the values are uncertain.

First, we observe that we can assume that the welfare-maximizing strategies of the bidders are threshold strategies. Again, we show that for each message chosen by bidder $i$, the welfare is a linear function in $v_i$. To show this we should use a backward-induction argument: in the last message, the bidders will clearly use thresholds. Therefore, in previous stages the welfare (as a function of $v_i$ fixing the strategies of all the other bidders) is a linear combination of linear functions which is itself a linear function. The maximum over linear function is a piecewise linear function and the thresholds will be its crossing points.

**Theorem 6.1.** *Let $h$ be an $n$-bidder sequential mechanism with a communication requirement $m$. Then, there exists a* simultaneous *mechanism $g$ that achieves, with dominant strategies, at least the same expected welfare as $h$, with a communication requirement smaller than $nm$.*

*Proof.* Consider an $n$-bidder mechanism $h$ with a Bayesian-Nash equilibrium, and with communication requirement $m$ (for simplicity, assume $n$ divides $m$, i.e., each bidder sends $\frac{m}{n}$ bits). There exists a profile $s = (s_1, ..., s_n)$ of threshold strategies that achieves the optimal welfare in $h$. First, we give an upper bound for the total number of thresholds each bidder uses in the game. For a bidder $i$, let $\alpha_1^i, ..., \alpha_{k_i}^i$ be the (positive) sizes of the $k_i$ messages she sends in $h$. Let $\gamma_j^i$ $(1 \leq j \leq k_i)$ be the number of bits that were sent by all the bidders (including $i$), *before* bidder $i$ sends his $j$th message. When choosing a message of size $\alpha_j^i$, the bidder uses up to $2^{\alpha_j^i} - 1$ thresholds. In each stage, every bidder can use a different set of thresholds, for every possible history of the game. Thus, for sending her $j$th message she can use up to $2^{\gamma_j^i} \left( 2^{\alpha_j^i} - 1 \right)$ different thresholds. Summing up, bidder $i$ uses at most $T(i) = \sum_{j=1}^{k_i} 2^{\gamma_j^i} \left( 2^{\alpha_j^i} - 1 \right)$ thresholds. Now, assume, w.l.o.g., that the bidders are numbered according to the order they send their *last* messages (i.e., $\gamma_{k_1}^1 > \gamma_{k_2}^2 > ... > \gamma_{k_n}^n$). Recall that the total number of bits sent by the bidders is $m$. When sending the last message, bidder 1 thus uses $2^{m - \alpha_{k_1}^1} \left( 2^{\alpha_{k_1}^1} - 1 \right) < 2^m$ different thresholds. Since each messages have a non-zero size, bidder 2 will have at most $2^{m-1-\alpha_{k_2}^2} \left( 2^{\alpha_{k_2}^2} - 1 \right) < 2^{m-1}$ different thresholds for the last stage. Similarly, every bidder $i$ can use at most $2^{m-i+1}$ thresholds for his last message. But therefore, for her before-last message bidder $i$ uses at most $2^{m-i-1}$ different thresholds (the worst case occurs when one bidder sends one bit between bidder $i$'s 2 last messages). It follows that the maximal number of different thresholds for bidder $i$ is:

$$
\begin{aligned}
T(i) \;=\; \sum_{j=1}^{k_i} 2^{\gamma_j^i} \left( 2^{\alpha_j^i} - 1 \right) \quad &<\quad 2^{m-i+1} + 2^{m-i-1} + \sum_{j=1}^{k_i-2} 2^{\gamma_j^i} \left( 2^{\alpha_j^i} - 1 \right) \\
<\quad 2^{m-i+1} + 2^{m-i-1} + \sum_{j=1}^{m-i-2} 2^j \quad &<\quad 2^{m-i+1} + 2^{m-i-1} + 2^{m-i-1} \quad <\quad 2^{m-i+2}
\end{aligned}
$$

---

24. Note that in sequential mechanisms the bidders must be informed about the bits the other bidders sent (we do not take this into account in our analysis), so the total gain in communication can be very mild.

Now, let $g$ be a simultaneous mechanisms in which each bidder simply "informs" the mechanism between which of the thresholds he uses in $h$ his valuation lies. Clearly, for every set of valuations of the bidders, this allocation in $g$ and $h$ is identical. Due to the inequality above, $m - i + 2$ bits suffice for bidder $i$ to express this number. We conclude that the number of bits sent by all the bidders in $g$ is smaller than: $\sum_{i=1}^n (m - i + 2) = nm - \frac{n(n-3)}{2}$.

Finally, we mention that we can set the allocation scheme and the payment scheme in $g$ such that the threshold-strategies based on the thresholds in $s$ will be an equilibrium and the expected welfare will not decrease. As shown in Section 3, we turn this mechanism to be monotone by allocating the item deterministically to the bidder with the highest expected value, in each combination of bids. A dominant-strategy equilibrium follows.

This analysis holds for any order and sizes of the bidder messages, even when they depend on the history of the messages, since counting the number of thresholds can be still done in the same way. □

## 7. Future Work

This paper concerns single-item auctions that are severely limited in their ability to elicit information from the bidders: only a few possible bids are available for each player although each player may have a continuum of types. We give a comprehensive analysis of such auctions, and present welfare- and revenue-maximizing mechanisms under these restrictions. We asymptotically analyze the losses in these optimal mechanisms compared to auctions with unrestricted communication, and we also compare them to auctions where the bidders' messages are sent sequentially.

We leave several questions open. The most obvious problem is the exact characterization of the optimal mechanisms for an arbitrary number of players and an arbitrary number of possible bids. This paper fully characterized the optimal 2-bidder $k$-bid auctions and the optimal $n$-bidder 2-bid auctions, but only presented *asymptotically* optimal results for the general case of $n$ players and $k$ bids. Also, future work may provide an asymptotic analysis of the welfare- and the revenue loss as a function of both $k$ and $n$ (we provided a separate asymptotic analysis by each of these variables).

An additional interesting question is regarding the gain from allocating the bits of communication non-uniformly among the agents. While in simple domains (like 2-bidder simultaneous auctions) uniform distribution of the communication seems to be the best option, this would be unclear, and probably untrue, in more general settings. In addition, it seems that the concepts and methods presented in this work extend to more general frameworks, like general single-parameter mechanism-design settings and mechanism design with interdependent values (some extensions are given in the recent work by Blumrosen & Feldman, 2006).

Finally, this work presented a partial study of *sequential* auctions with communication restrictions. This kind of auctions captures many reasonable real-life settings, and seems to be analytically challenging. We did not present a characterization of the optimal sequential mechanisms in this paper, nor a direct comparison of simultaneous and sequential mechanisms with the same communication requirement. Future work may also compare prior-aware sequential mechanisms and "detail-free" sequential mechanisms (a similar comparison for simultaneous mechanisms showed that detail-free mechanism can only achieve trivial results). Another possible extension would be to take an integrated approach and study settings with partially-known priors.

**Acknowledgment.** We thank Ron Lavi, Daniel Lehmann, Ahuva Mua'lem and Motty Perry for helpful discussions. We also thank several anonymous referees for valuable remarks, suggestions and insights. The first two authors were supported by grants from the Israeli Academy of Sciences. The third author was supported by the National Science Foundation.

## Appendix A. Missing proofs

In this section we present formal proofs for some of the results given in the body of the paper.

## A.1 Optimality of Threshold Strategies

### Proof of Claim 3.2 in Theorem 3.1:

*Proof.* Given a vector of strategies $s^*$ which achieve optimal welfare in $g$ (i.e., $\max_{\widetilde{s} \in \times_{i=1}^k \varphi_{k_i}} w(g, \widetilde{s})$ ), we will show that for every player $i$ we can modify $s_i^*$ to be a threshold strategy, and the welfare will not decrease.

Assume $s_i^*$ is not a threshold strategy. Therefore, there must be $\alpha, \beta, \gamma \in [\underline{a}, \overline{b}]$, $\alpha < \beta < \gamma$ such that $s_i^*(\alpha) = s_i^*(\gamma) = m$ but $s_i^*(\beta) \neq m$ (where $m$ is some bid of player $i$). We will show that a strategy vector $s$ identical to $s^*$, except that for every such $\beta$ $s_i(\beta) = m$, we have that $w(g, s) \geq w(g, s^*)$.

Denote the probability that all players except $i$ bids $b_{-i}$ as $Pr(b_{-i})$. Thus, the expected welfare from a game $g$ given that bidder $i$ with valuation $v_i$ bids $m$ and that the other players use strategies $s_{-i}^*$ is:

$$\sum_{b_{-i}} Pr(b_{-i}) \left( a_i(m, b_{-i}) \cdot v_i + \sum_{j \neq i} a_j(m, b_{-i}) \cdot E(v_j \, | \, s_j^*(v_j) = b_j) \right)$$

Note that this expected welfare is a linear function of $v_i$, and we denote it by $h(m) \cdot v_i + t(m)$ (the constants $h(m)$ and $t(m)$ depend on the bid $m$).

We know that $s^*$ achieve optimal welfare in $g$ and that $s_i^*(\alpha) = m$. Therefore, there is no other bid $l$ such that if $s_i^*(\alpha) = l$, the expected welfare will increase, i.e.:

$$\forall l \neq m \quad h(m) \cdot \alpha + t(m) \geq h(l) \cdot \alpha + t(l) \tag{10}$$

Similarly, because $s_i^*(\gamma) = m$:

$$\forall l \neq m \quad h(m) \cdot \gamma + t(m) \geq h(l) \cdot \gamma + t(l) \tag{11}$$

Because $\beta$ is a convex combination of $\alpha$ and $\gamma$, and due to Equations 10 and 11:

$$\forall l \neq m \quad h(m) \cdot \beta + t(m) \geq h(l) \cdot \beta + t(l)$$

Thus, the expected welfare for player $i$, given $v_i = \beta$, is maximal when she bids $m$. Therefore, when modifying $s_i^*$ such that $s_i^*(\beta) = m$ the total expected welfare will not decrease. We can repeat this process until $s_i^*$ becomes a threshold strategy.[25] $\qquad \square$

## A.2 Optimal Mechanisms Use All the Possible Bids

**Lemma A.1.** $w_{2,(k,k)}^{opt} > w_{2,(k-1,k)}^{opt}$ *for every $k > 1$.*

*Proof.* Let $g \in G_{2,(k-1,k)}$ be a deterministic, monotone mechanism that achieves the optimal welfare with threshold strategies based on the vectors $(x, y)$. Each row in $g$ is of the form $[A, ..., A, B, ..., B]$, and let $l_i \in \{0, ..., k\}$ be the first index in row $i$ in which $B$ wins. We will modify $g$ to $\widetilde{g} \in G_{2,(k,k)}$ by adding some missing row, and change the threshold strategy $x$ to $\widetilde{x} \in \Re^{k+1}$, such that the welfare strictly improves. We assume, w.l.o.g., that the thresholds are unique (i.e., $0 < x_1 < ... < x_{k-1} < 1$, $0 < y_1 < ... < y_{k-1} < 1$).

*Case* 1. The row $[B, ..., B]$ is in the game's matrix.

Let $x'_1 = \frac{E_{v_B}(v_B | 0 \leq v_B \leq y_1)}{2}$, and let $\widetilde{x} = (0, x'_1, x_1, x_2, ..., x_{k-2}, 1)$. We will create a new game $\widetilde{g}$ by adding the line $[B, ..., B]$ as the first line. It is easy to see that the allocation in $g$ and $\widetilde{g}$ is identical in all rows except the new one. When $v_A \in [0, x'_1]$ and $v_B \in [0, y_1]$ $\widetilde{g}$ allocates the item to B where $g$ allocated the item to A. The distribution functions are always positive, hence this will occur with a positive probability. Since $E(v_A | v_A \in [0, x'_1]) < x'_1 < E_{v_B}(v_B | 0 \leq v_B \leq y_1)$, the expected welfare has strictly increased. For higher bids of bidder B, the allocation in the first row will clearly be efficient now, therefore no welfare loss was incurred.

*Case* 2. The row $[B, ..., B]$ does not appear in $g$'s game matrix.

Due to the monotonicity, $g$ must have two rows $i$ and $i+1$ and two columns $j$ and $j+1$ such that we allocate the item to $B$ when the bids are $(i, j), (i, j+1)$ and to $A$ when the bids are $(i+1, j), (i+1, j+1)$. We will create a mechanism $\widetilde{g}$ by adding a row $i'$ identical to row $i+1$ except that $B$ wins in index $j+1$. The new threshold is constructed as follows:

**If $\mathbf{E(v_B | y_j \leq v_B \leq y_{j+1})} < \mathbf{x_{i+1}}$:**
Let $x'_{i+1} = E(v_B | y_j \leq v_B \leq y_{j+1})$, and let $\widetilde{x} = (0, x_1, ..., x_i, x'_i, x_{i+1}, ..., 1)$. As in previous cases, the welfare in all entries hasn't changed, except a strictly positive improvement in the $(i', j)$ entry.

**If $\mathbf{E(v_B | y_j \leq v_B \leq y_{j+1})} \geq \mathbf{x_{i+1}}$:**
Let $x'_i = E(v_B | y_{j+1} \leq v_B \leq y_{j+2})$ and let $\widetilde{x} = (0, x_1, ..., x_i, x'_i, x_{i+1}, ..., 1)$. We show that since $g$ is efficient $x_{i+1} < x'_i < x_{i+2}$: First, $E(v_B | y_{j+1} \leq v_B \leq y_{j+2}) > E(v_B | y_j \leq v_B \leq y_{j+1}) \geq x_{i+1}$; Also, since $A$ wins for the bids $(i+1, j+1)$, we have $E(v_B | y_{j+1} \leq v_B \leq y_{j+2}) \leq E(v_A | x_{i+1} \leq v_A \leq x_{i+2}) < x_{i+2}$. It follows that the expected welfare has strictly increased in the entry $(i', j+1)$, and has not decreased in all other entries. □

## A.3 Optimal Symmetric 1-bit Mechanisms

Following are the optimal 1-bit 2-bidder mechanisms assuming independent uniform distribution for all values. The socially-efficient symmetric 1-bit mechanism achieves an expected welfare of 0.625 compared to 0.648 that is achieved in an asymmetric 1-bit mechanism and $2/3$ that is achieved with unrestricted communication. Similarly, the revenue-maximizing symmetric 1-bit mechanism below achieves an expected profit of 0.385 compared to 0.39 with 1-bit symmetric mechanisms and $5/12 = 0.417$ with unrestricted communication (second-price auction with a reserve price).

The following mechanism achieves the optimal welfare among all the symmetric 1-bit mechanisms:

|   | 0 | 1 |
|---|---|---|
| 0 | w.p. $\frac{1}{2} A$ wins, pays 0<br>w.p. $\frac{1}{2} B$ wins, pays 0 | $B$ wins and pays $\frac{1}{4}$ |
| 1 | $A$ wins and pays $\frac{1}{4}$ | w.p. $\frac{1}{2} A$ wins, pays $\frac{1}{2}$<br>w.p. $\frac{1}{2} B$ wins, pays $\frac{1}{2}$ |

Proving the social efficiency of the mechanism can be done by the following idea: First note that a symmetric, efficient mechanism will clearly allocate the item to the player that bids 1 when the other player bids 0, and allocate with equal probabilities of $\frac{1}{2}$ when the bids are equal. With threshold strategies $(x, y)$ the expected welfare is:

$w(x, y) = x \cdot y \cdot \left(\frac{1}{2} \cdot \frac{x}{2} + \frac{1}{2} \cdot \frac{y}{2}\right) + x \cdot (1 - y) \cdot \frac{(1+y)}{2} + (1 - x) \cdot y \cdot \frac{(1+x)}{2} + (1 - x) \cdot (1 - y) \cdot \left(\frac{1}{2} \cdot \frac{(1+x)}{2a} + \frac{1}{2} \cdot \frac{(1+y)}{2}\right)$

Maximum is achieved when $(x, y) = (\frac{1}{2}, \frac{1}{2})$.

The mechanism below is the revenue-maximizing mechanism:

|   | 0 | 1 |
|---|---|---|
| 0 | No allocation | $B$ wins and pays $\frac{1}{\sqrt{3}}$ |
| 1 | $A$ wins, pays $\frac{1}{\sqrt{3}}$ | w.p. $\frac{1}{2}$   $A$ wins, pays $\frac{1}{\sqrt{3}}$ <br> w.p. $\frac{1}{2}$   $B$ wins, pays $\frac{1}{\sqrt{3}}$ |

The idea behind the optimality of the above mechanism over all the 1-bit symmetric mechanisms: in the profit-maximizing symmetric mechanism if a player bids 0 and the other bids 1, the latter wins and pays $x$. When both players bid 1, they will pay $\overline{x}$ with equal probabilities. It is easy to see that under the ex-post IR assumption, $x = \overline{x}$. The expected profit is thus: $r(x) = x(1-x)x + (1-x)xx + (1-x)(1-x)(\frac{1}{2}x + \frac{1}{2}x)$. Maximum is achieved $(x \in [0, 1])$ when $x = \frac{1}{\sqrt{3}}$.